\documentclass[aps,prl,10pt,twocolumn,footinbib,superscriptaddress,amsmath,amsfonts,amssymb,floatfix,notitlepage]{revtex4-2} 

\usepackage{CJK}
\usepackage{physics} 
\usepackage{soul} 
\usepackage[caption=false,subrefformat=parens,labelformat=parens]{subfig}
\usepackage{color} 
\usepackage[table,dvipsnames]{xcolor}
\usepackage[colorlinks,linktocpage,bookmarks=false]{hyperref}
\usepackage[normalem]{ulem} 
\usepackage[percent]{overpic}
\usepackage[T1]{fontenc}
\usepackage{graphicx}

\newcommand\myshade{85}
\definecolor{myrulecolor}{RGB}{150,20,0}
\colorlet{mylinkcolor}{violet}
\colorlet{mycitecolor}{YellowOrange}
\colorlet{myurlcolor}{Aquamarine}
\hypersetup{
	linkcolor  = myurlcolor!\myshade!black,
	citecolor  = mycitecolor!\myshade!black,
	urlcolor   =  myrulecolor!\myshade!black
}

\graphicspath{{FIG/}}
\usepackage{multirow} 
\usepackage{booktabs} 

\AtBeginDocument{
	\heavyrulewidth=.08em
	\lightrulewidth=.05em
	\cmidrulewidth=.03em
	\belowrulesep=.65ex
	\belowbottomsep=0pt
	\aboverulesep=.4ex
	\abovetopsep=0pt
	\cmidrulesep=\doublerulesep
	\cmidrulekern=.5em
	\defaultaddspace=.5em
}

\renewcommand\[{\begin{equation}}
\renewcommand\]{\end{equation}}

 \newcommand{\prlsection}[1]{\textbf{#1}\ ---\ }

\makeindex

\begin{document} 
	\begin{CJK*}{UTF8}{gbsn} 
		\title{Fragmented Spin Liquid and Shadow Pinch Points  in Dipole-Octupole Pyrochlore  Spin Systems}
        \author{Daniel Lozano-G\'omez} 
        \email{daniel.lozano-gomezg@tu-dresden.de}
        \affiliation{Institut f\"ur Theoretische Physik and W\"urzburg-Dresden Cluster of Excellence ct.qmat, Technische Universit\"at Dresden, 01062 Dresden, Germany}	
        \author{Han Yan (闫寒)} 
        \email{hanyan@issp.u-tokyo.ac.jp}
        \affiliation{Institute for Solid State Physics, The University of Tokyo,  Kashiwa, Chiba 277-8581, Japan} 
\date{\today}
\begin{abstract}
We study the classical dipole-octupole pyrochlore spin systems in an external magnetic field along the \([111]\) direction. We identify an intermediate fragmented spin liquid phase that precedes full spin saturation, characterized by the coexistence of three phenomena: U(1) spin liquid on the Kagome planes, spontaneous $\mathbb{Z}_2$ symmetry breaking, and partial spin polarization from explicit symmetry breaking. We show that even without dipolar interactions, the dipolar components can form a spin liquid driven by the octupolar spin liquid. The physics manifests itself experimentally as shadow pinch points: low-intensity pinch points underlying the strong Bragg peaks. We discuss how these discoveries are directly relevant to various experiments on dipole-octupole pyrochlore materials, including neutron scattering, magnetization, and magnetostriction. 
\end{abstract}
\maketitle
\end{CJK*}

\prlsection{Introduction}
Frustrated magnetic systems on the pyrochlore lattice \cite{Balents2010,Gardner-RMP,Hallas-AnnRevCMP,Rau2019ARCMP,Smith2025-AnnRevCMP} have long served as fertile ground for the exploration of exotic spin phases~\cite{Wen2007,Balents2010,Springer_frust,ZhouRevModPhys.89.025003,savaryQuantumSpinLiquids2016,Gingras_2014,Knolle_Moessner_2019,moessner2021book}, including classical 
~\cite{Yan2024a,Yan2024b,Fang2024PhysRevB,
Harris1997PhysRevLett,Bramwell-Science,castelnovoSpinIceFractionalization2012,udagawaSpinIce2021,HenleyARCMP,Gingras_2014,Taillefumier2017PhysRevX,lozano_2024chiral,chung_gingras_2023_2form} and quantum spin liquids~\cite{Balents2010,savaryQuantumSpinLiquids2016,Gingras_2014,Knolle_Moessner_2019,Gresista_QPLSL,Molavian2007,Onoda_quantummeling-PRL,Lee2012PhysRevB,shannon2012PRL,Huang2018PhysRevLett,Benton_DO_2020,Pace2021PhysRevLett,hosoiUncoveringFootprintsDipolarOctupolar2022,DesrochersPRL,sanders2024experimentallytunableqeddipolaroctupolar}, spin fragmentation~\cite{Brooks_fragmentation,Petit_fragmentation_2016}, and order-by-disorder phenomena~\cite{savary2012,ZhitomirskyPRL2012,Noculak_HDM_2023,Hickey_2024arxiv,francini_higher_rank_spin_nematics_2025}. 
In particular, rare-earth pyrochlores in which one spin component behaves as a conventional Ising dipole while the orthogonal components transform as magnetic octupoles -- the so-called dipole-octupole (DO) systems \cite{Huang2014PhysRevLett} -- have recently attracted intense interest, both  
experimentally~\cite{Gao_Chen_Tam_Huang_Sasmal_Adroja_Ye_Cao_Sala_Stone_2019,Sibille2020NatPhys,
poree2023fractional,poree2023dipolaroctupolar,EvansPhysRevX.12.021015,Smith2023PhysRevB,poreeCrystalfieldStatesDefect2022,bhardwajSleuthingOutExotic2022,PhysRevB.111.155137,PhysRevB.106.094425,
YahnePhysRevX.14.011005,Smith2025-AnnRevCMP,Gao2025NatPhy} and theoretically~\cite{PhysRevB.95.041106,PhysRevResearch.2.013334,PhysRevResearch.2.013066,PhysRevResearch.5.033169,PhysRevB.110.L081110,zhao2025classicalsymmetryenrichedtopological,Benton_DO_2020,sanders2024experimentallytunableqeddipolaroctupolar,DesrochersPRL,desrochersSymmetryFractionalizationGauge2023,zhou2025globalphasediagramcebased}. For the classical model in zero external magnetic field, a dominant antiferromagnetic octupolar exchange leads to an extensively degenerate ground-state manifold analogous to classical spin ice (CSI), but with the ``2-in-2-out'' rule enforced in the octupolar sector. Recent works have studied the effect of an applied external magnetic field in the extensively degenerate ground-state manifold~\cite{PhysRevB.111.155137,PhysRevB.106.094425,poree2023dipolaroctupolar,sanders2024experimentallytunableqeddipolaroctupolar,PatriPhysRevResearch,zhou2024arxiv,zhou2025globalphasediagramcebased}. When a magnetic field is applied along the global $[111]$ direction -- which is one of the most important and common experiment on the relevant single crystal materials --  conventional Ising spin ices exhibit Kagome spin liquid behavior with 2-in-1-out constraints on the Kagome planes and fully polarized triangular layers~\cite{Moessner_Sondhi_2003,Isakov_Moessner_2004}. In DO pyrochlores, however, the interplay between octupolar correlations and Zeeman coupling of the dipoles gives rise to qualitatively new physics beyond those of canonical spin ice~\cite{sanders2024experimentallytunableqeddipolaroctupolar,zhou2025globalphasediagramcebased,PatriPhysRevResearch}. Understanding this interplay is not only a crucial theoretical challenge but also of direct relevance to imminent experiments on candidate materials.

\begin{figure}[ht!]
    \centering
    \begin{overpic}[width=\columnwidth]{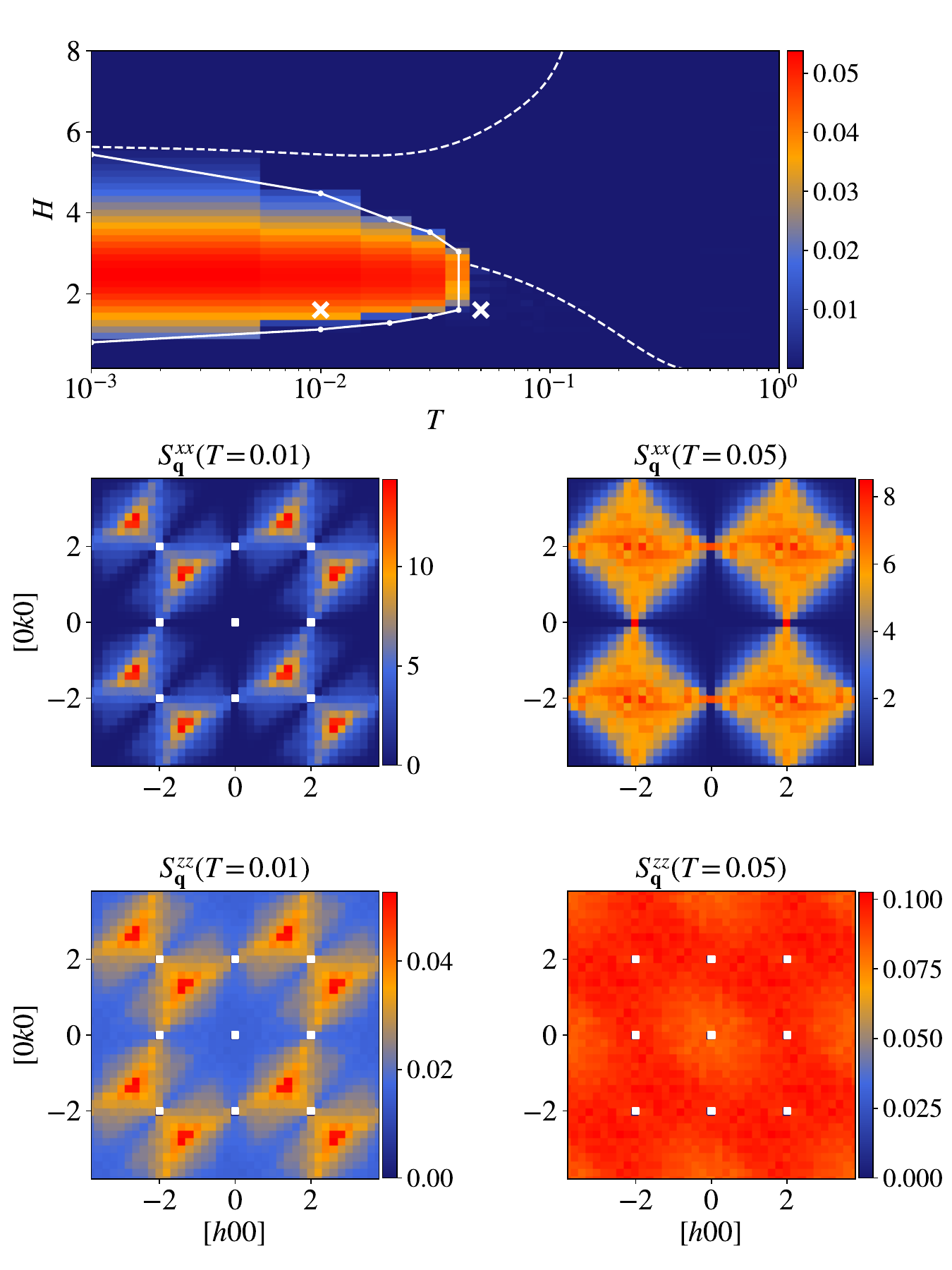}
        \put(30,72){\fontsize{7}{16} \textcolor{white}{ $\rm Pyrochlore\ SL$}}
        \put(13,78){ \textcolor{white}{$\substack{\rm Fragmented\ \\ \rm Kagome\ SL} $}}
         \put(18,92){\textcolor{white}{$\substack{\rm Polarized \\ \rm 3-in-1-out }$}}
        \put(40,85){\textcolor{white}{$ \substack{\rm Correlated \\ \rm PM}  $ }}
        \put(3,100){(a)}
        \put(3,64){(b)}
        \put(42,64){(c)}
        \put(3,32){(d)}
        \put(42,32){(e)}
        \put(61,98){$A_2^x$}
    \end{overpic}
    \caption{(a) Phase diagram of the Dipolar-Octupolar Hamiltonian using the interaction parameters $\{J_x,J_y,J_z,J_{xz}\}=\{1, 0,  0, 0\}$ in an external  $[111]$  field. Here, the color bar parametrizes the value of the $A_2^x$ parameter, where a non-zero value characterizes the $\mathbb{Z}_2$ symmetry-broken phase. The white full and dashed lines mark the symmetry-breaking transition and crossover temperature, respectively. (b) and (c) illustrate the  $S^x$-spin structure factors for the two temperature and external magnetic field points marked by the white X markers in panel (a) where the system is in the $\mathbb{Z}_2$ symmetry-broken phase and the U(1) SI phase, respectively. (d) and (e) illustrate the  $S^z$-spin structure factors for the same temperature and external fields used in panels (b) and (c), respectively. The white dots in the spin-structure-factor panel mark the position of magnetic Bragg peaks, which have been subtracted from the correlation  function. 
    }
     \label{fig:phase_diagram}
\end{figure}
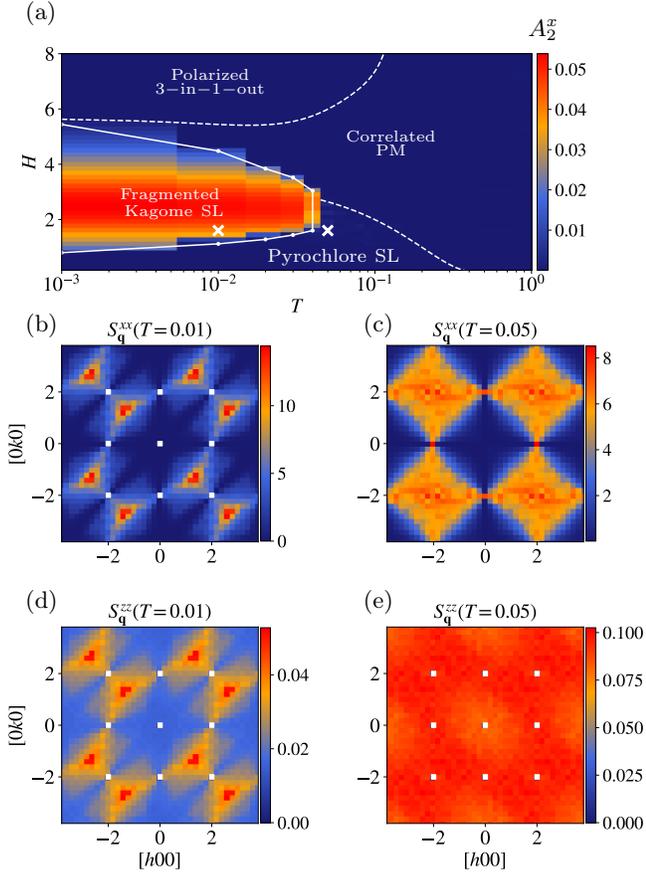

In this work, we investigate the effect of an external applied magnetic field along the global $[111]$  direction on a classical DO Hamiltonian, focusing on the regime in which a single octupolar coupling dominates. Through a combination of single-tetrahedron analysis and large-scale Monte Carlo simulations, we identify low-temperature ``fragmented spin liquid'' phase at intermediate magnetic fields that precedes full saturation, see Fig.~\ref{fig:phase_diagram}(a). This phase is characterized by (i) a fragmented U(1) classical spin liquid confined to the Kagome planes, (ii) spontaneous breaking of a global $\mathbb{Z}_2$ symmetry associated with the octupolar order parameter, and (iii) partial dipolar polarization enforced by the field. 
Microscopically, the octupolar degrees of freedom stabilize algebraic correlations, manifesting as ``shadow pinch points'' that underlie strong Bragg peaks in both octupolar and dipolar structure factors, with the latter visible to neutron-scattering experiments,  see Fig.~\ref{fig:phase_diagram}(b)-(e). We further demonstrate that these signatures are observable in cerium-based DO pyrochlores such as $\rm Ce_2Hf_2 O_7$ under accessible experimental conditions.

\prlsection{The Dipolar-Octupolar Hamiltonian}
Our starting point is the DO pyrochlore spin model~\cite{Huang2014PhysRevLett} coupled to the external magnetic field, described by the Hamiltonian
\begin{eqnarray}
\mathcal{H} &=& \sum_{\langle ij\rangle} J_x S^{x}_iS^{x}_j +J_y S^{y}_iS^{y}_j+J_z S^{z}_iS^{z}_j \nonumber\\
&& + J_{xz}( S^{x}_iS^{z}_j+S^{z}_iS^{x}_j) - g_z \sum_i (\vb*{H}\cdot \hat{\vb*{z}_i}) S^z_i,    \label{eq:DO_Hamiltonian}
\end{eqnarray}
where $S^\alpha_i$ are the spin components in the local basis, $\vb*{H} $ is the external magnetic field, and $\hat{\vb*{z}_i}$ is the unit vector of the local $[111]$ axis on site $i$, for the explicit basis vectors we refer the reader to the Supplementary Material (SM). Here, $S^z_i$ is defined along the local $[111]$ axis and acts as a conventional dipole component. On the other hand, the $S^y_i$ components of the spins transform as octupoles under the symmetries of the lattice and do not couple to an external field linearly~\cite{Rau2019ARCMP}. 
$S^x_i$ is also an octupole, but has symmetry properties identical to those of $S^z$. In the Hamiltonian in Eq.~\eqref{eq:DO_Hamiltonian} we only consider a Zeeman coupling of the external magnetic field to the $S_i^z$ components of the spins as it has been found that the coupling to the other spin component $S^x_i$ is usually negligible in experiments for pertinent materials, and is therefore set to zero~\cite{Smith2025-AnnRevCMP}.
We are interested in the parameter regime where $J_x$ dominates~\footnote{For practical purposes, and in many aspects of experimental observables, this is equivalent to having a dominant  $J_y$ coupling.}  and for an external magnetic field $\vb*{H}$ applied along the global  $[111]$  direction, one of the most common experimental setups ~\cite{Smith2025-AnnRevCMP,Smith2023PhysRevB,poree2023fractional,smith2025arxiv,YahnePhysRevX.14.011005,Sibille2020NatPhys,Bhardwaj2022NPJQM}. 
We begin  our investigation by taking  $J_x = g_z = 1$ and $J_{y} = J_z = 0 $ for a representative, minimal model capturing the core physics, and then discuss experimentally estimated parameters. 

\prlsection{Classical ground states}
Let us now understand the zero-temperature ground-state physics at an intermediate $H \equiv |\vb*{H}|$.
First, when the antiferromagnetic coupling $J_{x}$ dominates and $H  =0$, the classical model shows an extensively degenerate ground-state manifold identical to a classical spin ice (CSI) for the component $S^{x}$~\cite{Benton_DO_2020}. 
In conventional CSI ($J_z$ dominant), the ground states are spin configurations such that two spins point inward and two spins point outward on each tetrahedron, known as the 2-in-2-out ice rule. In the case of a dominant $J_x$ in the DO spin model, the spin configurations in the ground-state manifold also follow a ``2-in-2-out'' rule, but now rotated to the local $\hat{\vb*{x}}$ axis. 
On the other hand, when $H$ is strong enough to saturate the system, the ground state is an ordered $\hat{\vb*{z}}$ axis 3-in-1-out state.
This is different from the canonical Ising CSI, in which the spins are always constraint to be aligned along the local $\hat{\vb*{z}}$ axis at any $H$. 

We now focus on the physics during the magnetization process with intermediate $H$. Since the  Hamiltonian has a vanishing $J_y$ interaction, the spins in the ground states always lie in the local $xz$ plane of each sublattice site. The application of an external field breaks time-reversal symmetry. However, since the field only couples to the local $S^z$ degrees of freedom, the model preserves a $\mathbb{Z}_2$ symmetry on the $S_i^x$ components, i.e. $S_i^x \leftrightarrow -S_i^x$, as well as one in the $S_i^y$ components. Therefore, every ground state is unique up to an overall $-1$ sign in  $S_i^x$.

\begin{figure}[t!]
    \centering
    \begin{overpic}[width=\columnwidth]{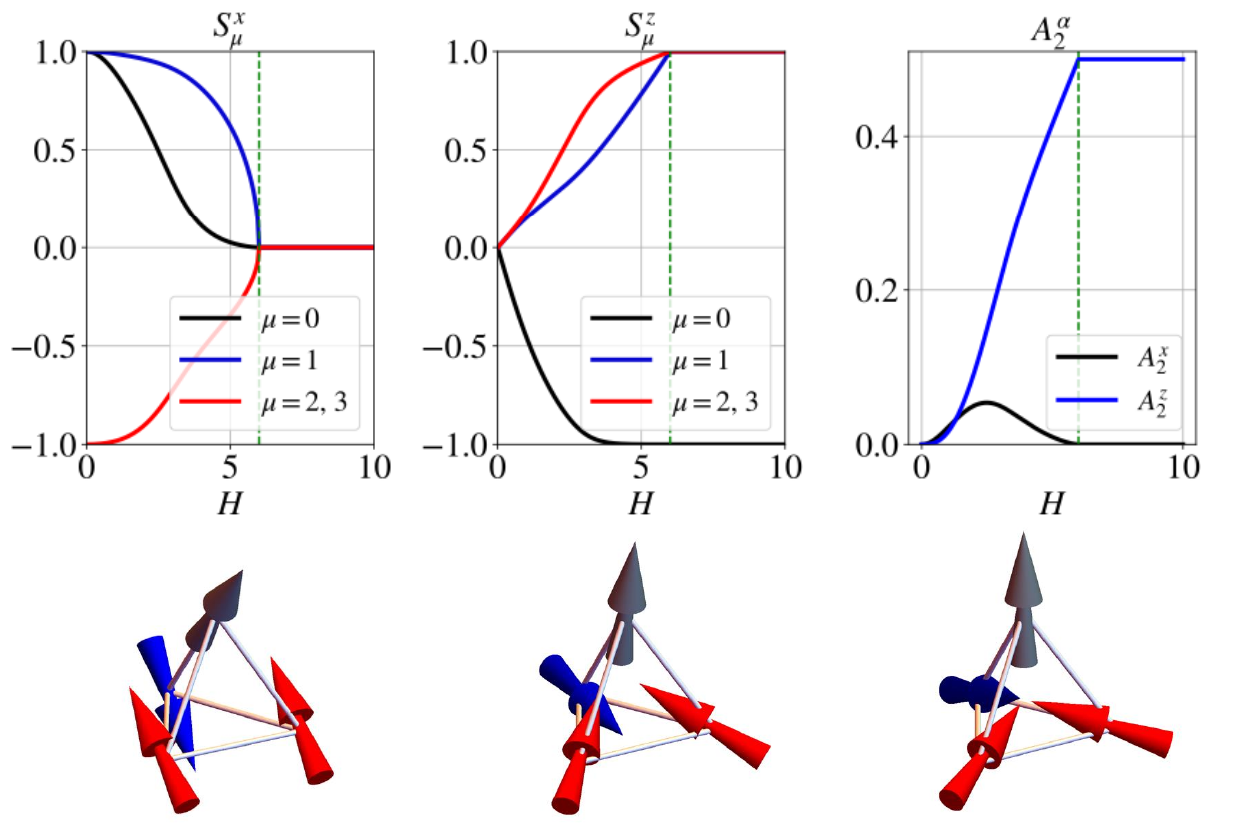}	
        \put(3,66){(a)}
        \put(35,66){(b)}
        \put(68,66){(c)}
        \put(3,19){(d)}
        \put(35,19){(e)}
        \put(68,19){(f)}
    \end{overpic}
    
    \caption{\label{fig:Single_tet} Evolution of ground-state $S^x_\mu$ (a),  $S^z_\mu$ (b), and  $A^\alpha_2$ (c) in a single tetrahedron as a function of field magnitude $H$. Here the vertical dashed green line indicates the field value where the system enters the fully saturated phase in which all tetrahedra adopt a 3-in-1-out configuration. (d), (e), and (f) illustrate the spin configuration in a single tetrahedron for the  $H=1.5$, $H=3$, and $H=7$, respectively. The colors in panels (d), (e), and (f) correspond to the spin sublattice shown in panels (a) and (b). }
\end{figure}
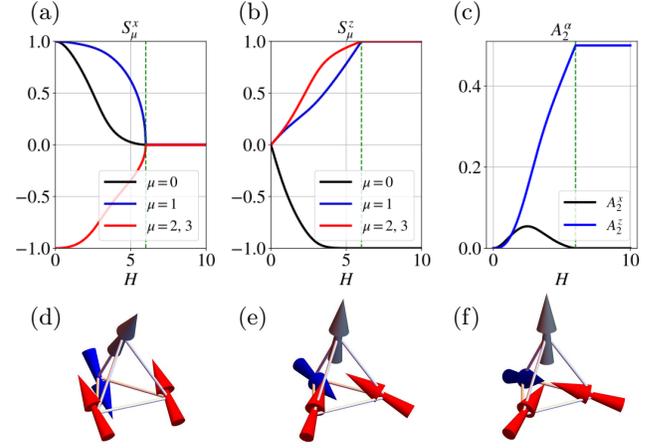

As $H$ increases, the competition between the octupolar ($S^x$) and the dipolar ($S^z$) components caused by the Zeeman interaction becomes a determining factor of the ground state configuration. The competition is  between spins tending to align along the local $\hat{\vb*{z}}$ axis to reduce the Zeeman energy and forming a $S^x$ 2-in-2-out states to minimize the    antiferromagnetic interactions.
As a consequence, the ground states at intermediate $H$ fields before saturation fulfill the following rules: the four spins on a tetrahedron take the values
\begin{eqnarray}
    \vb*{S}_0 &=& (X_0, 0, Z_0), \\
\{\vb*{S}_1,\vb*{S}_2,\vb*{S}_3 \} & = & \{ (X_1,0, Z_1),(X_2,0, Z_2),(X_2, 0, Z_2) \} \nonumber \\
&&\text{or the other two permutations}
\end{eqnarray}
and also those states from applying the global symmetry  $S^x \rightarrow - S^x$. 
The exact values of $X_i$ and $Z_i$  are shown in Fig.~\ref{fig:Single_tet}(a,b), respectively, along with their net local moments $A_2^\alpha = \sum_{i=0}^3 S_i^\alpha$ in panel (c). For the definition of the net local moments we refer the reader to the SM. We note that all the spins located in the triangular layers are fixed to a single value (modulo the $\mathbb{Z}_2$ symmetry of $S^x \rightarrow - S^x$), progressively aligning with the magnetic field (Fig.~\ref{fig:Single_tet}(b), black line), resulting in a long-range-ordered state. In contrast, for the Kagome layers, six spin configurations on every triangle are obtained, where one spin aligns along the local $(X_1,0 , Z_1)$ direction, while the other two  along the local $(X_2 , 0, Z_2)$ direction. 
We denote this as the ``2-red-1-blue'' rule  on the Kagome plane, which in turn defines a 
U(1) classical spin liquid on the Kagome plane. 
In Fig.~\ref{fig:Single_tet} (d-e), some examples of the single-tetrahedron spin configuration obtained at different external fields are shown. 
Note that since $X_1 \ne X_2$ for finite $H$, this is \textit{not} identical to Kagome ice, which allows both 2-up-1-down and 1-up-2-down. This distinction makes a qualitative difference in the nature of the two spin liquids: our Kagome spin liquid is a U(1) spin liquid, whereas the plain Kagome ice is a paramagnet.
Furthermore, we note that the $\mathbb{Z}_2$ symmetry of the
$S^x$ components gives rise to two sets of ground states, differing by the sign of the local net $x$ moment $A_2^x$. The two sets of ground states, related by $S^x \leftrightarrow -S^x$, cannot be connected to each other without energy cost. That is, there is an energetically costly  domain wall of the associated $\mathbb{Z}_2$ order parameter $A_2^x$.

We have thus discovered an exotic zero-temperature phase for the classical  DO pyrochlore system under magnetic field: 
it features the coexistence of layers of Kagome-plane U(1) spin liquid,  a global $\mathbb{Z}_2$ symmetry in 3D, which must be spontaneously broken at low temperatures (as we also expose with our numerical simulations in the next section) and its associated order parameter taking finite value, as well as magnetization due to explicit symmetry breaking. The coexistence of a  spontaneous symmetry-breaking order \emph{and} a classical spin liquid in the intermediate field phase grants this phase the name of a ``fragmented spin liquid''.
In comparison the conventional Ising CSI has no such no fragmentation.
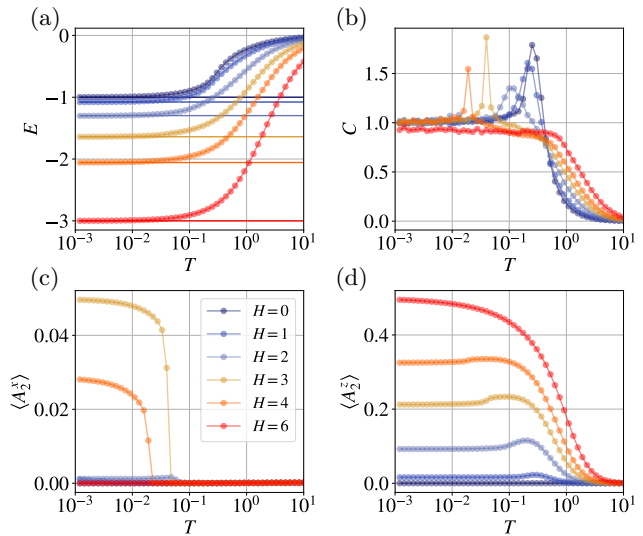
\begin{figure}[t!]
    \centering
    \begin{overpic}[width=\columnwidth]{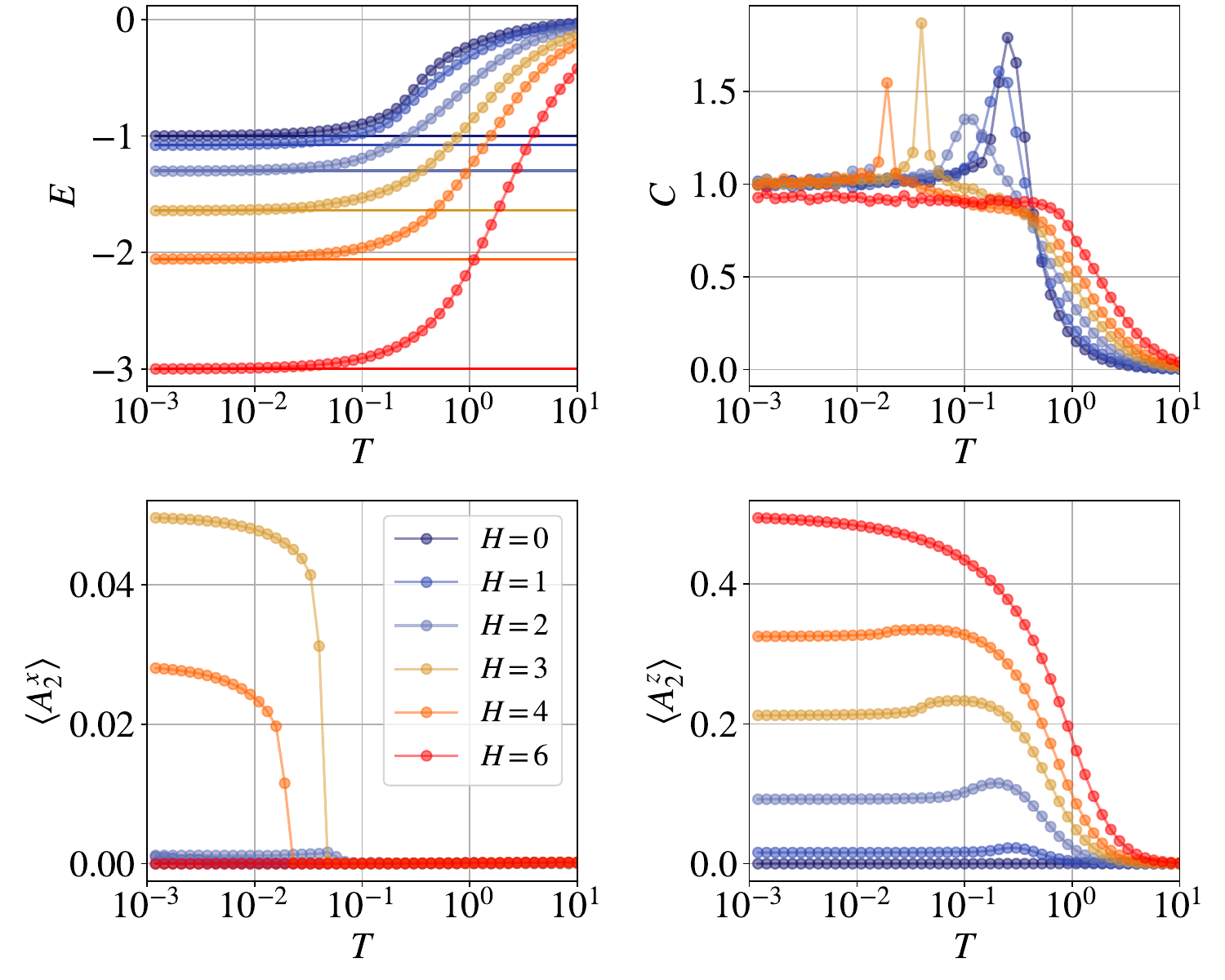}
    \put(5,80){(a)}
    \put(52,80){(b)}
    \put(5,40){(c)}
    \put(52,40){(d)}
    \end{overpic}
    \caption{\label{fig:thermodynamics}
    Thermodynamic quantities obtained via classical Monte Carlo for distinct fixed external fields along the global  $[111]$  direction as a function of temperature. Here, the horizontal lines in the upper-leftmost panel mark the single-tetrahedron energy predicted in the  $T\to 0$ limit.
    } 
\end{figure}

\prlsection{Finite Temperature Results}
To further characterize the Hamiltonian in Eq.~\eqref{eq:DO_Hamiltonian} under the application of an external magnetic field, we perform a classical Monte Carlo (CMC) simulation. In what follows, we first consider a system whose interaction parameters adopt the values $(J_x,J_y,J_z,J_{xz})=(1,0,0,0)$, 
and then study more general parameters relevant to the experimentally discovered materials. For more details regarding the CMC simulations, we refer the reader to the SM.

Figure~\ref{fig:thermodynamics} illustrates the internal energy, specific heat, and thermodynamic average of the net $\alpha$ moment order parameter $\langle A_2^\alpha\rangle$ per site as a function of temperature $T$ obtained for a variety of external magnetic field amplitudes. Figure~\ref{fig:thermodynamics}(a) illustrates the energy per tetrahedron evolution as a function of temperature, where the horizontal lines towards which the energy curves plateau mark the ground-state energy as predicted by the single-tetrahedron analysis. Figure~\ref{fig:thermodynamics}(b) illustrates the temperature evolution of the specific heat. For the case of $H=0$, the specific heat develops a Schottky-like peak, similar to the one observed in conventional spin ice, signaling the depopulation of emergent magnetic monopoles in the system~\footnote{In contrast to the classical Ising spin ice, the specific heat plateaus to $1$ in the limit $T\to 0$, as opposed to $0$ in the classical Ising model. This reflects the fact that the spin components considered are continuous and that the low-temperature  fluctuations about the ground-state manifold are quadratic~\cite{ChalkerHoldsworthKagomePhysRevLett.68.855,lozanogomez2023arxiv}.}. As the magnetic field is increased, the Schottky peak shifts to lower temperatures, reflecting the competing effect between the Zeeman energy and the antiferromagnetic interactions. At slightly higher fields within the temperature window studied, a sharp peak in the specific heat is observed. At temperatures where these specific heat peaks occur, our simulations detect an increase of the thermally averaged local $A_2^x$ parameter.
The observation of a sharp peak in the specific heat together with the evolution of the $A_2^x$ parameter provides evidence for a $\mathbb{Z}_2$ symmetry-breaking transition in the $S^x$ components of the spins.  
Lastly,   Fig.~\ref{fig:thermodynamics}(d)   illustrates the thermally averaged $A_2^z$ parameter. This last parameter plateaus to $0.5$ in the low-temperature-high-field regime where all tetrahedra adopt a 3-in-1-out configuration, i.e. the maximally polarized configuration. 
We note that the value to which the $A_2^\alpha$ parameter plateaus in the low-temperature regime is accurately captured by the single-tetrahedron analysis, see Fig.~\ref{fig:Single_tet}. Altogether, the phase diagram as a function of external magnetic field and temperature is shown in Fig.~\ref{fig:phase_diagram}(a). For a further discussion on the $\mathbb{Z}_2$ symmetry breaking we refer the reader to the SM. 

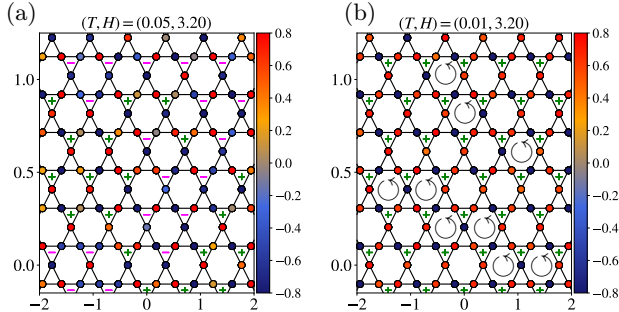
\begin{figure}[ht!]
    \centering
    \begin{overpic}[width=1.\columnwidth]{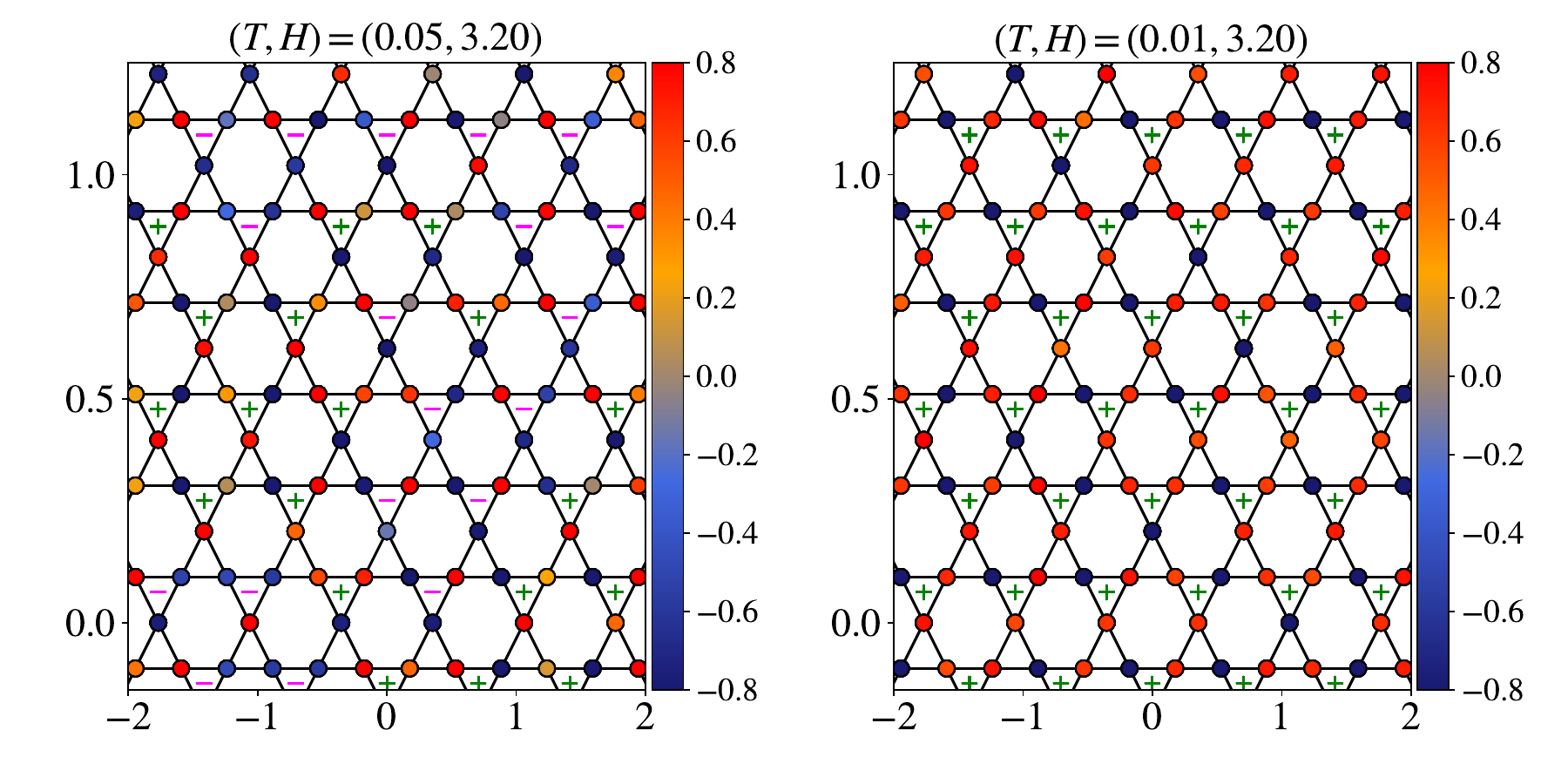}	
        \put(3,48){(a)}
        \put(55,48){(b)}
        \put(67.5,38.5){\fontsize{12}{10} $\circlearrowleft$}
        \put(70.5,32.5){\fontsize{12}{10} $\circlearrowleft$}
        \put(79.2,26.5){\fontsize{12}{10} $\circlearrowleft$}
        \put(58.7,20.6){\fontsize{12}{10} $\circlearrowleft$}
        \put(64.5,20.6){\fontsize{12}{10} $\circlearrowleft$}
         \put(67.5,14.8){\fontsize{12}{10} $\circlearrowleft$}
         \put(73.5,14.8){\fontsize{12}{10} $\circlearrowleft$}
          \put(76.4,8.9){\fontsize{12}{10} $\circlearrowleft$}
         \put(82.3,8.9){\fontsize{12}{10} $\circlearrowleft$}
    \end{overpic}
    \caption{\label{fig:Kagome_cut}
 Snapshots of a single Kagome plane in the pyrochlore lattice for two different temperatures where the sites are colored by the $S^x_\mu$ component. The $+$ and $-$ signs label the local $A_2^x$ sign measured in each the down triangles. It is worth noting that for the $T=0.01$ only positive moments are observed in the down triangles, signaling the $\mathbb{Z}_2$ symmetry breaking in the system. In panel (b) the black circular arrows mark some of the possible hexagonal loops which can be permuted.
    } 
\end{figure}

\prlsection{Fragmented Spin Liquid Phase and Shadow Pinch Points}
As illustrated in the previous section, in the $\mathbb{Z}_2$ symmetry-broken phase, the spin components $S^x$ adopt a long-range ordered configuration in the triangular planes while remaining in a disordered configuration on the Kagome planes. This is shown in  Fig.~\ref{fig:Kagome_cut}, where a snapshot of a randomly sampled Kagome plane is shown above (a) and below (b) the symmetry-breaking transition, and the $\pm$ is the sign of the local $A_2^x$  parameter.  
This symmetry-broken phase can be understood as a fragmented spin liquid described by an effective electric field, defined microscopically as $\vb*{E}_i = S_i^x \hat{\vb*{z}_i}$.
It can be separated into two components by a Helmholtz-Hodge decomposition~\cite{Brooks_fragmentation,Benton_origins_2016,Petit_fragmentation_2016}, a divergence-free and a divergence-full component, i.e. $\vb*{E} = \vb*{E}_\parallel + \vb*{E}_\perp$. In this phase, the $\vb*{E}$ satisfies an energetically imposed Gauss' law, $\vb*{\nabla}  \cdot \vb*{E}= \vb*{\nabla}  \cdot \vb*{E}_\parallel = A_2^x$,
which describes the partial order in the system. The remaining divergence-free component of the field acts as a spin liquid, fulfilling a vanishing Gauss' law constraint, $\vb*{\nabla} \cdot  \vb*{E}_\perp=0$, resulting in algebraic correlations between the components of this field~\cite{Isakov_correlations_2004,Yan2024a}. The decomposition of the field $\vb*{E}$ is reflected in the spin component  $S^x$ correlation function in the $\mathbb{Z}_2$ symmetry-broken phase in the $[hk0]$ reciprocal plane, see Fig.~\ref{fig:phase_diagram}(b). In this correlation function, Bragg peaks and  pinch point features are observed simultaneously, where the Bragg peaks are associated with the partial order of the system (described by the $\vb*{E}_\parallel$ component) while the pinch points expose algebraic correlations taking place in the disordered Kagome planes (described by the $\vb*{E}_\perp$ component).

\prlsection{Experimental Signatures of the Fragmented Spin Liquid Phase} The features manifested in the $S^x$-component spin correlation function,
whose octupolar nature means it is not easily  accessible in experiments at the wave vector range studied in neutron-scattering experiments~\cite{poree2023dipolaroctupolar}. 
However, as explained in the classical ground-state analysis, the dipolar components  $S^z$ of the spins, whose correlations are more accessible, display a similar behavior in the $\mathbb{Z}_2$ symmetry-broken phase ground-state to   the  $S^x$ components, see Fig.~\ref{fig:phase_diagram}(d). In this phase, each triangle on the Kagome plane has a fixed 2-red-1-blue rule  for both the $S^x$ and $S^z$, and this gives rise to a U(1) spin liquid for the fragmented $S^z$ components on these planes  too. As such, coexistence of Bragg peaks and sharp pinch points is expected in the elastic neutron-scattering experiments.
When the $\mathbb{Z}_2$ symmetry is recovered at higher temperatures, the U(1) topological physics on the Kagome plane is lost, and the $S^z$ pinch point disappears, see Fig.~\ref{fig:phase_diagram}(e). 
Note that the system can still be in a 3D pyrochlore U(1) spin liquid phase, as shown in Fig.~\ref{fig:phase_diagram}(a) and (c), so the pinch-point features observed in the $S^x$ components associated with the Kagome U(1) spin liquid disappear, but those associated with the pyrochlore U(1) spin liquid remain.

\begin{figure}[ht!]
    \centering
    \begin{overpic}[width=\columnwidth]{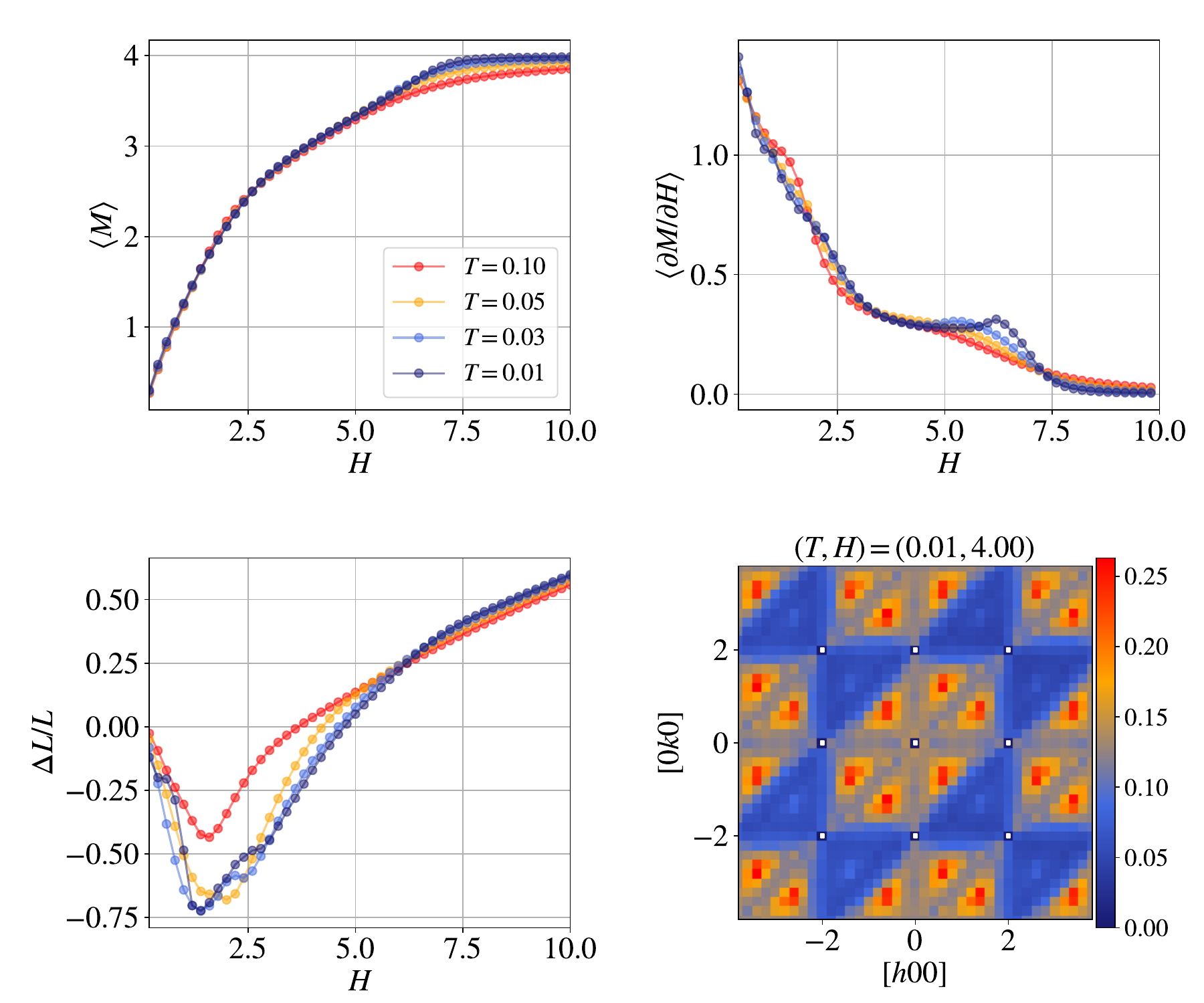}	
    \put(2,83){(a)}
    \put(50,83){(b)}
    \put(2,40){(c)}
    \put(50,40){(d)}
    \end{overpic}
    \caption{\label{fig:CeHfO}
Evolution of $z$-all-in-all-out $A_2^z$ (a), $\partial A_2^z/\partial H$ (b), and magnetostriction $\Delta L/L$ (c) as a function of external magnetic field for various temperatures. (d) Neutron structure factor in the $[hk0]$ plane. The quantities illustrated in this figure are obtained using the interaction parameters $\{J_x,J_y,J_z,J_{xz}\}=\{1.,0.478,  0.239, -0.0217\}$ corresponding to the parametrization for $\rm Ce_2 Hf_2 O_7$ obtained by Ref.~\cite{poree2023dipolaroctupolar}, where we have taken the $J_x$ coupling as the measure of energy.
   }
\end{figure}
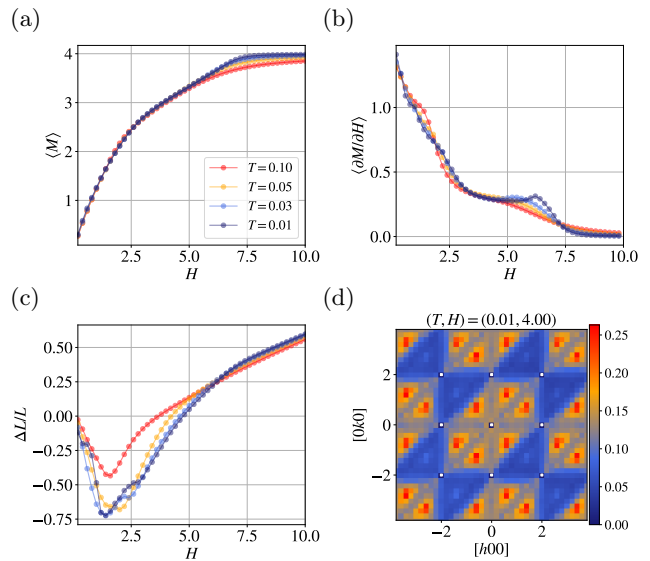

\prlsection{Application to Cerium compounds}
So far, we have discussed how the application of an external magnetic field along the global  $[111]$  direction results in a symmetry-breaking transition in which the low-temperature phase is a fragmented spin liquid. Although the analysis presented considers a Hamiltonian which has a single non-zero nearest-neighbor octupolar coupling, the fragmented spin liquid phase can also be realized for a more general set of parameters where one of the octupolar components is dominant and antiferromagnetic in nature. These conditions are fulfilled by some of the recently synthesized $\rm Ce$-based compounds, see Table~\ref{tab:materials}. Indeed, using the parameters for $\rm Ce_2 Hf_2 O_7$ obtained in Ref.~\cite{poree2023dipolaroctupolar}, we found that the similar evolution of the spin order parameter and symmetry-breaking transition, as those shown in Fig.~\ref{fig:Single_tet} and Fig.~\ref{fig:thermodynamics} respectively, are observed at low temperatures. As such, we expect that, at low temperatures and under the application of an external magnetic field along the  $[111]$  direction, $\rm Ce_2 Hf_2 O_7$ can support such a fragmented spin liquid phase. Experimental signatures  are expected to be observed not only in the dipolar ($z$-component) correlation functions, but also in the evolution of the magnetization, susceptibility, and magnetostriction as a function of the field at low temperatures. Figure~\ref{fig:CeHfO} illustrates the field dependence of the magnetization $\langle M\rangle$ parameter, the susceptibility  $\langle \partial M/\partial H\rangle$ associated with it, the magnetostriction $\Delta L/L$ along the [111] direction for various temperatures, and neutron structure factor in the $[hk0]$ scattering plane in the fragmented spin liquid phase, we refer the readers to the SM for the precise definition of these quantities. As the temperature is decreased, the $\langle M\rangle$ curves rise, and two distinct slopes can be identified as a function of magnetic field, see Fig.~\ref{fig:CeHfO}(a). These two characteristic slopes result in two distinctive features in the derivative $\langle\partial M/\partial H\rangle$; a sharp decrease of the susceptibility at low fields and a bump at higher fields, see Fig.~\ref{fig:CeHfO}(b). A similar (but inverted) behavior is observed in the magnetostriction where, at low temperatures,  a two-valley shape develops at low fields, succeeded by two increasing linear regimes at intermediate and higher fields, see Fig.~\ref{fig:CeHfO}(c). Lastly, Fig.~\ref{fig:CeHfO}(d) illustrates the neutron structure factor in the $[hk0]$ plane, where similar features to those shown in Fig.~\ref{fig:phase_diagram}(c) are observed.

\prlsection{Conclusion and discussion}
We have shown that classical DO pyrochlore magnets with dominant octupolar antiferromagnetic interaction in an external $[111]$ magnetic field host a novel fragmented spin liquid phase, in which emergent U(1) spin-liquid behavior on the Kagome planes coexists with $\mathbb{Z}_2$ symmetry breaking and partial dipolar polarization. This fragmentation yields a unique combination of Bragg peaks and pinch-point features (``shadow pinch points'') in both octupolar and dipolar spin correlation channels. 
Our results provide clear experimental predictions -- magnetization anomalies,  magnetostriction signatures, and characteristic structure-factor patterns -- for $\rm Ce_2 Hf_2 O_7$ and related compounds. 
More broadly, the interplay of multipolar exchange and external fields in DO systems offers a versatile platform for engineering and detecting hybrid ordered-liquid states in frustrated magnets. 

Beyond the classical model, it would be interesting to study whether the fragmented spin liquid phase is still stabilized when quantum effects are introduced. Recent works~\cite{sanders2024experimentallytunableqeddipolaroctupolar,zhou2025globalphasediagramcebased} found that, through the application of a small external magnetic field, the nature of the quantum spin ice can be tuned from the U$(1)_0$ or U$(1)_\pi$ to other flux-patterned spin liquid at zero temperature. 
Although our results are classical, they are relevant for spin-1/2 DO pyrochlore systems at finite temperatures. This is reflected in the reproduction of the magnetization curve of $\rm Ce_2 Hf_2 O_7$~\cite{chunjiong} (cf. Fig.~ \ref{fig:CeHfO_parameters_experiments} and SM). 
It remains an interesting question to connect the finite temperature, effectively classical physics with the zero-temperature quantum phase in magnetic field~\cite{sanders2024experimentallytunableqeddipolaroctupolar,zhou2025globalphasediagramcebased}, and explore how results in each regime help understanding each other.\\


\begin{acknowledgments}
\prlsection{Acknowledgments}
The authors greatly appreciate inspiring discussions with Minoru Yamashita,  Bruce D. Gaulin, Chun-Jiong Huang, Edwin Kermarrec,  Evan M. Smith, Peter Holdsworth, Matthias Vojta, Pedro Cons\^oli, and Michel J. P. Gingras. 
D. L.-G. acknowledges financial support from the DFG through the Hallwachs-R\"ontgen
Postdoc Program of the W\"urzburg-Dresden Cluster of Excellence on Complexity and
Topology in Quantum Matter -- \textit{ct.qmat} (EXC 2147, project-id 390858490) and
through SFB 1143 (project-id 247310070).
H.Y. acknowledges the 2024 Toyota Riken Scholar Program from the Toyota Physical 
and Chemical Research Institute, and the  Grant-in-Aid for Research Activity Start-up from Japan Society
for the Promotion of Science (Grant No. 24K22856).
\end{acknowledgments}
 
\bibliography{superliquid_shadow_pinch_points_ref.bib}


\clearpage
\setcounter{equation}{0}
\setcounter{figure}{0}
\setcounter{table}{0}
\makeatletter
\renewcommand{\theequation}{S\arabic{equation}}
\renewcommand{\thefigure}{S\arabic{figure}}
\renewcommand{\thetable}{S\arabic{table}}
\renewcommand{\bibnumfmt}[1]{[#1]}
\renewcommand{\citenumfont}[1]{#1}

\onecolumngrid
 
\begin{center}
	\Large{\textbf{Supplementary Materials for ``Fragmented Spin Liquid and Shadow Pinch Points  in Dipole-Octupole Pyrochlore  Spin Systems''}}
\end{center} 
 
\section{Local basis}
\begin{figure}[h!]
    \centering
    \begin{overpic}[width=0.3\columnwidth]{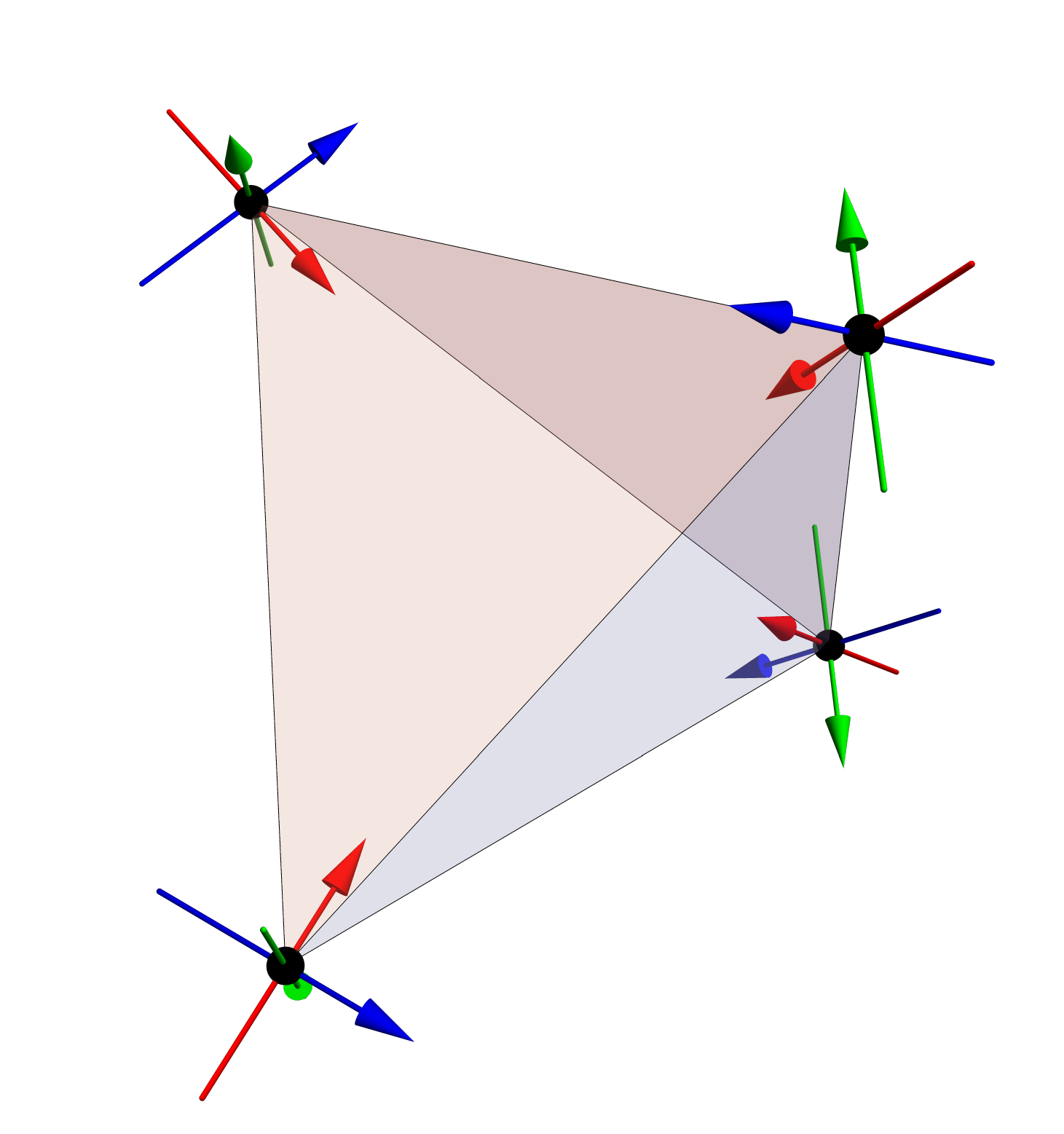}	
    \put(10,10){$\vb*{S}_0$}    
    \put(82,40){$\vb*{S}_1$}    
    \put(7,82){$\vb*{S}_3$}    
    \put(80,80){$\vb*{S}_2$}    
    \end{overpic}
    \caption{\label{fig:single_tet}
   Single tetrahedron local basis vectors. Here the blue, green, and red directions correspond to the local $\vb*{x}$, $\vb*{y}$, and $\vb*{z}$ directions, respectively.
    } 
\end{figure}

In this section, we provide the local $x$ and local $z$ directions for the spins in a single tetrahedron illustrated in Fig.~\ref{fig:single_tet}.  The remaining local $y$ directions are obtained via the cross-product of these two.
\begin{eqnarray}
    \vb*{x}_0&=\frac{1}{\sqrt{6}}(-2,1,1),\\
    \vb*{x}_1&=\frac{1}{\sqrt{6}}(-2,-1,-1),\\
     \vb*{x}_2&=\frac{1}{\sqrt{6}}(2,1,-1), \\  
     \vb*{x}_3&=\frac{1}{\sqrt{6}}(2,-1,1),
\end{eqnarray}
and 
\begin{eqnarray}
    \vb*{z}_0&=\frac{1}{\sqrt{3}}(1,1,1),\\
    \vb*{z}_1&=\frac{1}{\sqrt{3}}(1,-1,-1),\\
     \vb*{z}_2&=\frac{1}{\sqrt{3}}(-1,1,-1), \\  
     \vb*{z}_3&=\frac{1}{\sqrt{3}}(-1,-1,1).
\end{eqnarray}

\label{appendix:local_coordinates}
 
\section{Classical Monte Carlo Details}
\label{appendix:CMC_details}
For our simulations, we considered a system defined with an FCC unit cell composed of $4L^3$ spins with $L=10$. We thermalize the system, then implement a variety of local and cluster updates. For the local updates we implement a Gaussian spin-flip update~\cite{Alzate-Cardona_2019}, and an over-relaxation update~\cite{ZhitomirskyPRL2012,Creutz}, in addition to averaging up to $10$ independent CMC runs. For the cluster updates we use a generalized version of the loop algorithm developed in Ref.~\cite{lozanogomez2023arxiv} for all Cartesian components of the spins. Additionally, we developed an additional loop algorithm dubbed the permutation-loop algorithm. This algorithm identifies a closed loop of alternating red-blue spins in a randomly chosen Kagome plane and then permutes \emph{all} the spin components by assigning them to the next site in the loop. We note that, by construction, the permutation-loop algorithm does not change the energy of the new state. We note that the application of the generalized version of the loop algorithm and permutation-loop significantly improve the statistics of our CMC simulations. In particular, we note that, without the permutation-loop, the system would effectively freeze into a single configuration in the fragmented spin liquid phase. Indeed, both loop algorithms not only allow us to increase our statistics but also give us access to the distinct topological sectors in the spin ice phase and the fragmented spin liquid phase by identifying loops which wrap around the system.

We performed $8\times10^4$ thermalization sweeps and $ 2\times10^5$ measurement sweeps where we measured the energy, specific heat, and spin structure factors of the systems considered. We study the Hamiltonian in Eq.~\eqref{eq:DO_Hamiltonian} with an applied magnetic field along the global  $[111]$  direction for several magnitudes of the field $H$. We study this system by both varying the temperature at a fixed external magnetic field, and by varying the field at a fixed temperature.

\section{Thermodynamic quantities and correlation functions}
In this section, we provide the explicit expressions of the thermodynamic quantities measured in the CMC simulations, the $S^\alpha$-structure factors and the neutron structure factor presented in the main text. The $A^\alpha_2$ order parameter corresponds to the net moment of a tetrahedron along the local $\alpha$ component, 
\begin{eqnarray}
    A^\alpha_2&=&\frac{1}{4}\sum_{\rm i\in  tetra} S^{\alpha}_i.
\end{eqnarray}
For continuous spins, the $A^\alpha_2$ parameter yields values in the range $[-1,1]$, where $A^\alpha_2=0$ corresponds to a 2-in-2-out configuration axis, $A^\alpha_2=0.5$ ($A^\alpha_2=-0.5$) to a 3-in-1-out (1-in-3-out) configuration, and  $A^\alpha_2=1$ ( $A^\alpha_2=-1$) to the all-in-all-out (all-out-all-in) configuration, where the spins point along their local $\alpha$-axis. Another thermodynamic quantity of interest presented in the main text corresponds to the magnetostriction along the $[111]$ direction upon the application of a magnetic field, defined as \cite{PatriPhysRevResearch}
\begin{eqnarray}
\begin{aligned}
\left(\frac{\Delta L}{L}\right)= & \frac{H}{27 c_B}\left[\left(g_{10}+2 g_9\right)\left(3 \langle S_0^z \rangle-\langle S_1^z \rangle-\langle S_2^z \rangle-\langle S_3^z \rangle\right)+\left(g_4+2 g_3\right)\left(3 \langle S_0^x \rangle-\langle S_1^x \rangle-\langle S_2^x \rangle-\langle S_3^x \rangle\right)\right] \\
& +\frac{4H}{27 c_{44}} \left[\left(8 \sqrt{2} g_1-4 g_2\right)\left(\langle S_1^x \rangle+\langle S_2^x \rangle+\langle S_3^x \rangle\right)+\left(g_4-g_3\right)\left(9 \langle S_0^x \rangle+\langle S_1^x \rangle+\langle S_2^x \rangle+\langle S_3^x \rangle\right)\right] \\
& +\frac{4H}{27 c_{44}}\left[\left(8 \sqrt{2} g_7-4 g_8\right)\left(\langle S_1^z \rangle+\langle S_2^z \rangle+\langle S_3^z \rangle\right)+\left(g_{10}-g_9\right)\left(9 \langle S_0^z \rangle+\langle S_1^z \rangle+\langle S_2^z \rangle+\langle S_3^z \rangle\right)\right],
\end{aligned} \label{eq:magnetostriction}
\end{eqnarray}
where the coefficients $c_B = 1$, $c_{44}=1$, and $g_i$ for $i\in \{1,2,3,4,7,8,9,10\}$ take values~\cite{chunjiong}
\begin{align}
8 \sqrt{2} g_1-4g_2& =  0.723 \\
    g_3 &= -0.358 \\
g_4 &=  0.606 \\
8 \sqrt{2} g_7-4g_8& = -0.241 \\
g_9 &= 0.012 \\
g_{10}& = -0.024  
\end{align}

In the main text, we mainly discuss two types of correlation functions; the $S^\alpha$-structure factors and the neutron structure factor. These correlation functions are defined using the general correlation between spins which is given by the expression 
\begin{align}
      \mathcal{S}^{\alpha\gamma}_{\mu\nu}=\langle S_{\mu}^\alpha (\vb*{q}) S_{\nu}^\gamma(-\vb*{q})\rangle ,
\end{align}
where $\mu,\nu$ label the sublattices and $\alpha,\beta$ label the spin components.  
The $S^\alpha$-structure factors is defined as 
\begin{eqnarray}
\mathcal{S}^{\alpha\alpha}(\vb*{q})&=&\sum_{\mu,\nu}\langle S_{\mu}^\alpha (\vb*{q}) S_{\nu}^\alpha(-\vb*{q})\rangle,
\end{eqnarray}
which uniquely studies the correlation between the $\alpha$ components of the spins. In addition to  $S^\alpha$-structure factors, we have also computed the experimentally measurable unpolarized neutron structure factor, given by the expression
\begin{equation}
 \mathcal{S}_\perp(\vb*{q}) = \sum_{\alpha,\beta}\sum_{\mu,\nu}\left(\delta_{\alpha,\beta} -\hat{\vb*{q}}^\alpha \hat{\vb*{q}}^\beta\right)\langle m_{\mu}^\alpha (\vb*{q}) m_{\nu}^\beta(-\vb*{q})\rangle,\label{eq:unpolarized}
\end{equation}
which measures the correlation between the dipolar magnetic moments, $\vb*{m}_{i\mu}$, of the systems considered where the sub-index $i$ labels the primitive FCC vectors $\vb*{R}_i$ and $\mu$ denotes the sublattice sites, defined as 
\begin{eqnarray}
    m_{i\mu}^\alpha=\sum_{\beta } g_\mu^{\alpha\beta}S_{i\mu}^\beta,
\end{eqnarray}
where $g_\mu^{\alpha\beta}$ is the g-tensor, which in the local basis takes the form
\begin{eqnarray}
  g_\mu=\begin{pmatrix}
      g_{xy} & 0 &0 \\
      0 & g_{xy} & 0 \\
      0 & 0 & g_{zz} 
  \end{pmatrix} . 
\end{eqnarray} 
Since the spins in the local basis of the dipolar-octupolar Hamiltonian only have a $S^z_i$ dipolar component, we consider a $g$-tensor where $g_{xy}=0$. Consequently, for the dipolar-octupolar spin system the neutron structure factor simplifies to the expression
\begin{equation}
 \mathcal{S}_\perp(\vb*{q}) =\sum_{\alpha,\beta}\sum_{\mu,\nu}\left(\delta_{\alpha,\beta} -\hat{\vb*{q}}^\alpha \hat{\vb*{q}}^\beta\right) \vb*{z}_\mu^\alpha \vb*{z}_\mu^\beta \langle S_{\mu}^z (\vb*{q})   S_{\nu}^z(-\vb*{q})\rangle,\label{eq:Sunpolarized}
\end{equation}
where $\vb*{z}_\mu^\alpha$ corresponds to the $\alpha$ component of the local $z$ direction of the spin in the $\mu$ sublattice, and for simplicity we set $g_{zz}=1$.

\begin{figure*}[h!]
    \centering
   \includegraphics[width=\textwidth]{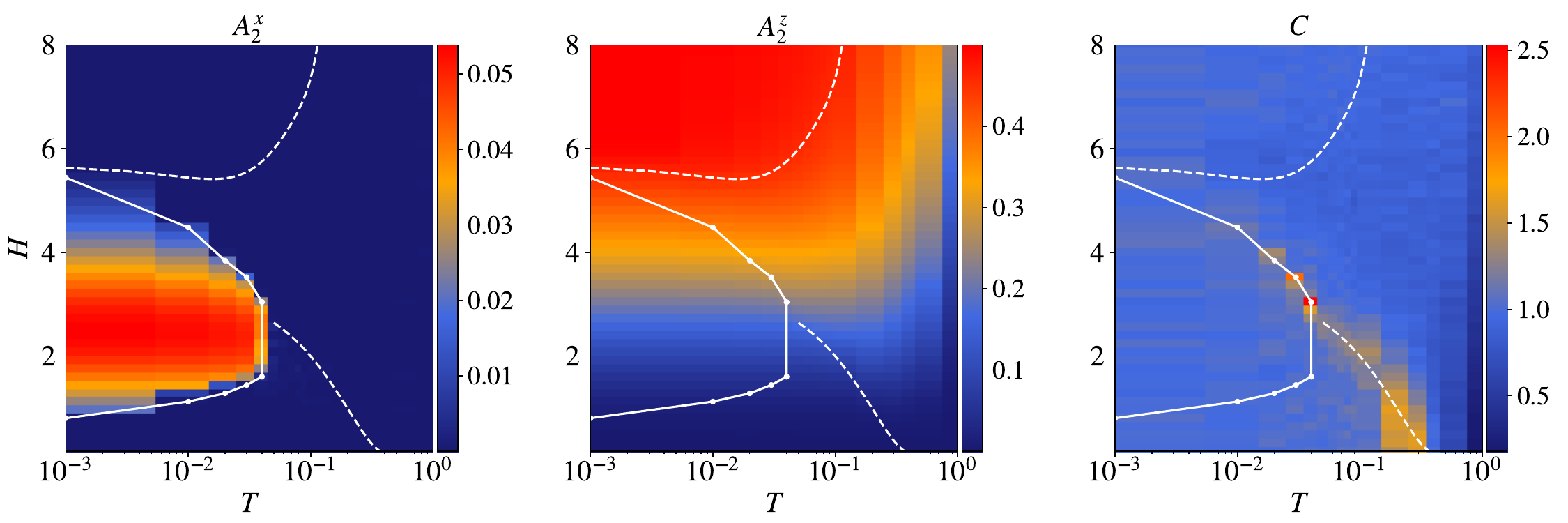}	
    \caption{\label{fig:phase_diagram_data}
    Thermodynamic  quantities obtained via classical Monte Carlo for distinct fixed temperatures as a function of external magnetic field for the set of parameters $\{J_x,J_y,J_z,J_{xz}\}=\{1,0,0,0\}$. Here the white dots mark the set of parameters for which the spin distributions and the structure factors are calculated.
    } 
\end{figure*}

\section{Extended Phase Diagram}
The phase diagram shown in Fig.~\ref{fig:phase_diagram}(a) of the main text is obtained by studying the temperature and field dependence of the $A_2^x$ parameter, the $A_2^z$ parameter, and specific heat; the $\mathbb{Z}_2$ symmetry-breaking transition is identified by the observation of a non-zero $A_2^x$ parameter, the crossover from the paramagnetic phase to the U(1) spin ice phase is identified as the temperature where the specific heat presents a maximum associated with the Schottky-like peak, and the crossover from the PM to the polarized state is identified as the temperature at which the $A_2^z$ reaches $90\%$ saturation of the maximal value (i.e. $A_2^z=0.45$). Additionally, the polarized 3-in-1-out phase is identified when the $A_2^z$ parameter plateaus to a value of approximately $0.5$. Figure~\ref{fig:phase_diagram_data} depicts the temperature and field dependence of the $A_2^x$, $A_2^z$, and specific heat $C$ measured via CMC.

\section{Detailed Evolution of the Spin Structure Factor}
\label{appendix:Evolution_structure_factors}
\begin{figure}[h!]
    \centering
   \includegraphics[width=0.8\columnwidth]{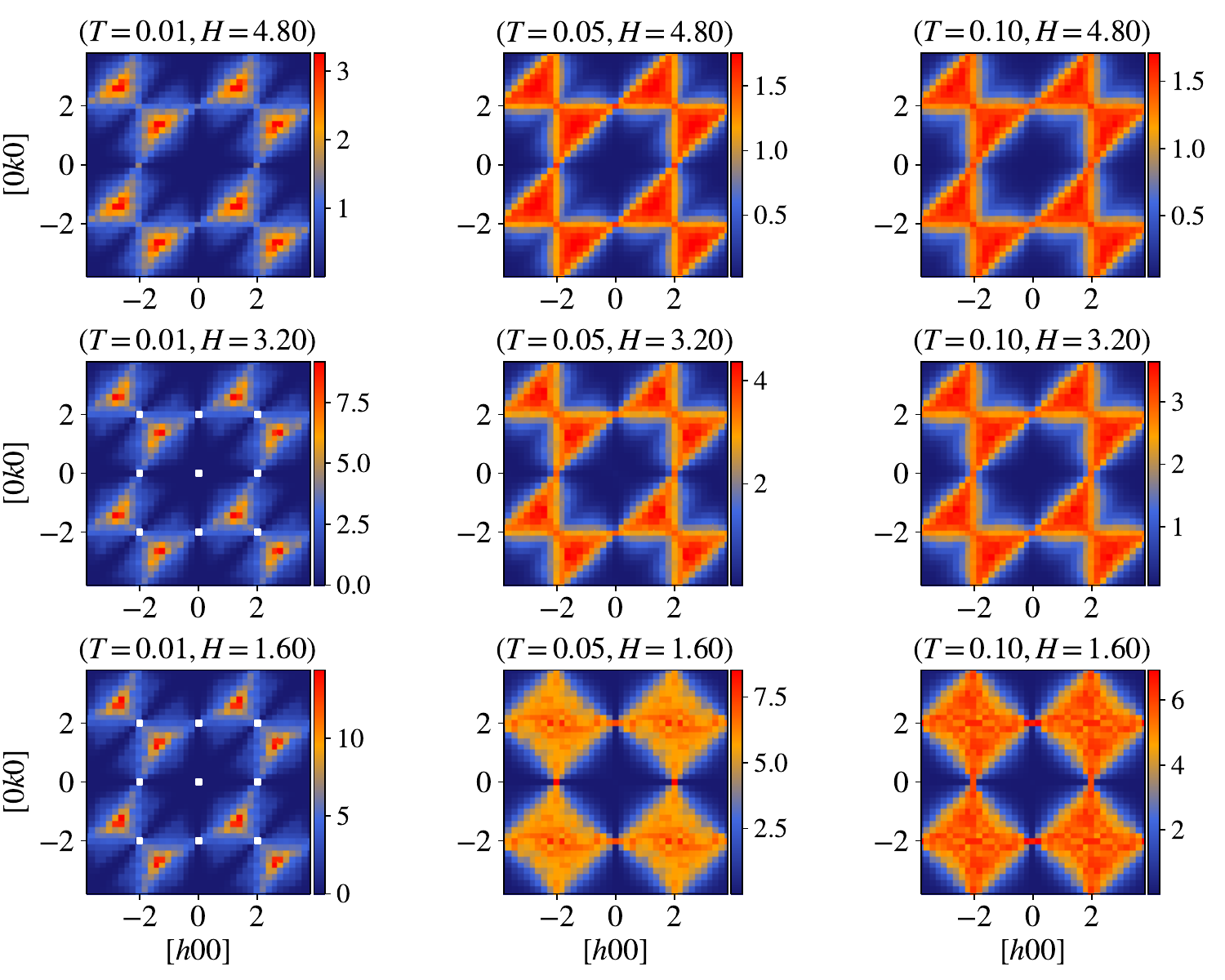}
    \caption{\label{fig:Structure_Sx_all}
    Temperature and field dependence of the $x$-component spin-spin correlation function in the $[hk0]$ scattering plane for a system with the interaction couplings $\{J_x,J_y,J_z,J_{xz}\}=\{1,0,0,0\}$. In this figure, white dots mark the points where magnetic Bragg peaks are observed. The Bragg peaks in these correlation functions have been subtracted in order to reveal the underlying diffuse scattering.
    } 
\end{figure}

In this section, we provide further analysis of the $x$- and $z$-component spin correlation functions discussed in the main text as a function of temperature and external magnetic field. Figure~\ref{fig:Structure_Sx_all} illustrates the $x$-component spin correlation function in the $[hk0]$ plane for three temperatures and magnetic field strengths. At low fields and intermediate temperatures, the system is in the octupolar U(1) spin ice phase, where the regular diamond-shaped spectrum and pinch points associated with a spin ice spin liquid are observed. As the magnitude of the field is increased, the spins in the triangular layers start to be pinned by the external magnetic field, while the Kagome layers remain disordered. Additionally, the diamond-shaped features become pinched in their center, leading to additional pinch points at $[200]$ and other symmetry-related points. As the temperature is decreased, the pinch points become sharper and, below the phase transition, magnetic Bragg peaks appear whose locations are marked by the white dots. As discussed in the main text, these Bragg peaks are associated with the spins in the triangular layers which adopt a uniform configuration throughout the lattice.

\begin{figure}[h!]
    \centering
   \includegraphics[width=0.8\columnwidth]{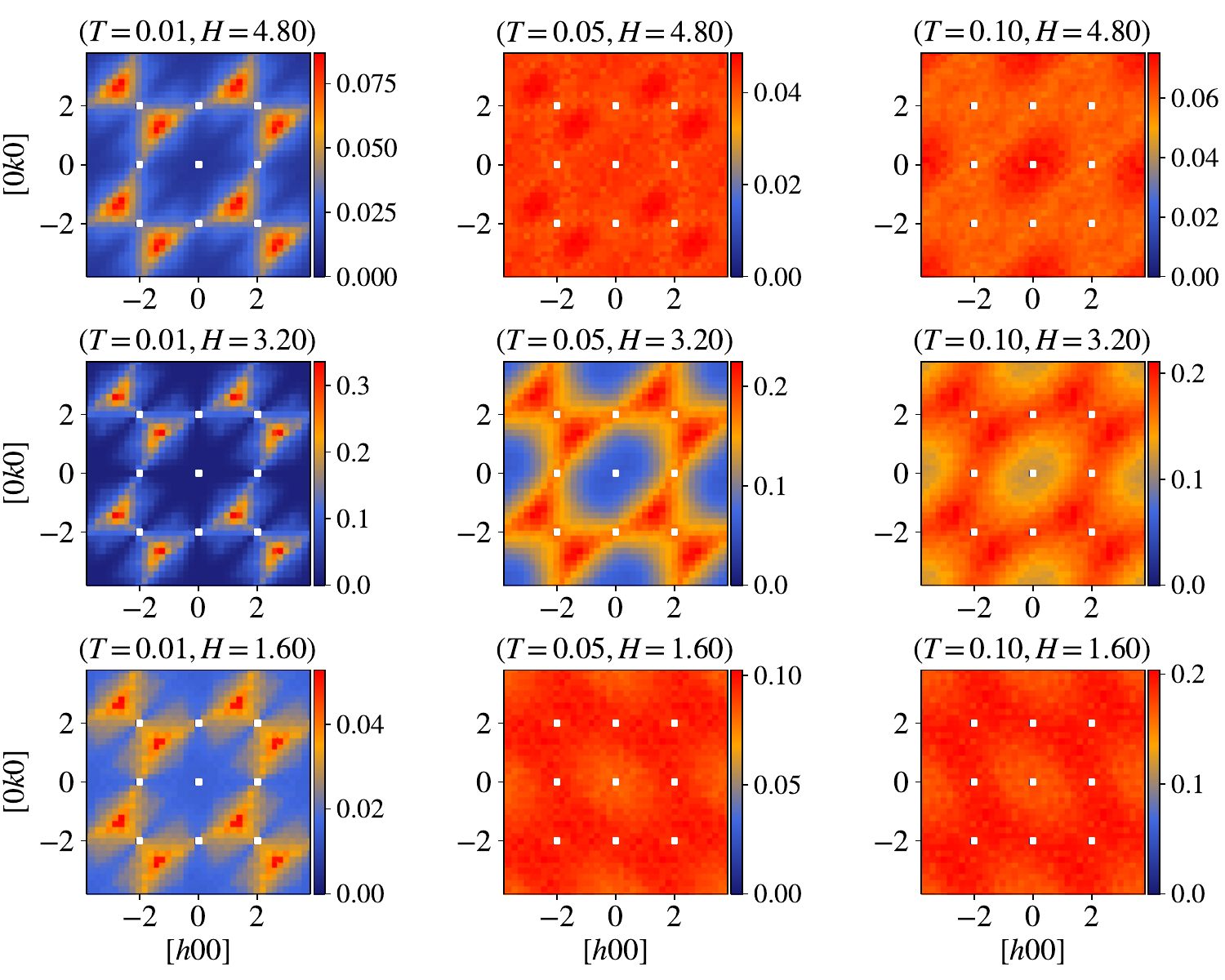}
    \caption{\label{fig:Structure_Sz_all}
    Temperature and field dependence of the $z$-component spin-spin correlation function in the $[hk0]$ scattering plane for a system with the interaction couplings $\{J_x,J_y,J_z,J_{xz}\}=\{1,0,0,0\}$. In this figure, white dots mark the points where magnetic Bragg peaks are observed. The Bragg peaks in these correlation functions have been subtracted in order to reveal the underlying diffuse scattering.
    } 
\end{figure}

Figure~\ref{fig:Structure_Sz_all} illustrates the $z$-component spin correlation function in the $[hk0]$ plane for the same temperatures and fields as those used in Fig.~\ref{fig:Structure_Sx_all}. In these correlation functions, the magnetic Bragg peaks have been subtracted in order to expose the underlying correlations in the system. These Bragg peaks marked by white dots reflect the overall tendency of the system to polarize  the spins along the global  $[111]$  direction. The diffuse scattering observed in these correlation functions exposes roughly uniformly distributed correlations for temperatures and magnetic field strengths distant from the $\mathbb{Z}_2$ symmetry-broken phase. In contrast, for temperatures and fields within or in the proximity of the $\mathbb{Z}_2$ symmetry-broken phase, pinch points are observed. As discussed in the main text, the observation of these anisotropic features is associated with the collapse of the system into one of the $\mathbb{Z}_2$ symmetry sectors.

\section{$\mathbb{Z}_2$ Symmetry Breaking}
As discussed in the main text, the $\mathbb{Z}_2$ symmetry breaking observed in the CMC simulations is associated with the selection of the  charge sector, either positive or negative, of the $A_2^x$ order parameter. In this section, we provide further analysis regarding this transition by further studying the spin configurations sampled in the simulation shown in Fig.~\ref{fig:Kagome_cut} in the main text and analyzing distribution of the spin components on the Kagome layers as a function of external magnetic field. Figure~\ref{fig:Kagome_cut} shows a section of a randomly selected Kagome plane above and below the symmetry-breaking transition, where the sites $i $ are colored according to their corresponding $S^x_i$ components. In this figure, each down triangle is assigned a ``$\pm$'' according to the sign of the local $A_2^x$ parameter. As shown in Fig.~\ref{fig:Kagome_cut}(a), at a temperature above the transition, the local $A_2^x$ parameter adopts both positive and negative values. In contrast, after the system is cooled below the transition, the $A_2^x$ parameter becomes homogeneous by selecting a unique sign of the local  $A_2^x$ parameter across the lattice, see Fig.~\ref{fig:Kagome_cut}(b). This selection reflects the $\mathbb{Z}_2$ symmetry breaking that has taken place. We note, however, that the observation of a symmetry-breaking transition does not imply that all the spin  degrees of freedom adopt an ordered configuration. Furthermore, by inspecting Fig.~\ref{fig:Kagome_cut}(b), one can identify a set of minimal hexagonal loops on the Kagome plane marked by black circular arrows, which can be permuted with no associated energy cost. The identification of the minimal loops and the associated ground-state degeneracy supports  a U(1) spin liquid.\\
\begin{figure}[h!]
    \centering
   \includegraphics[width=0.75\columnwidth]{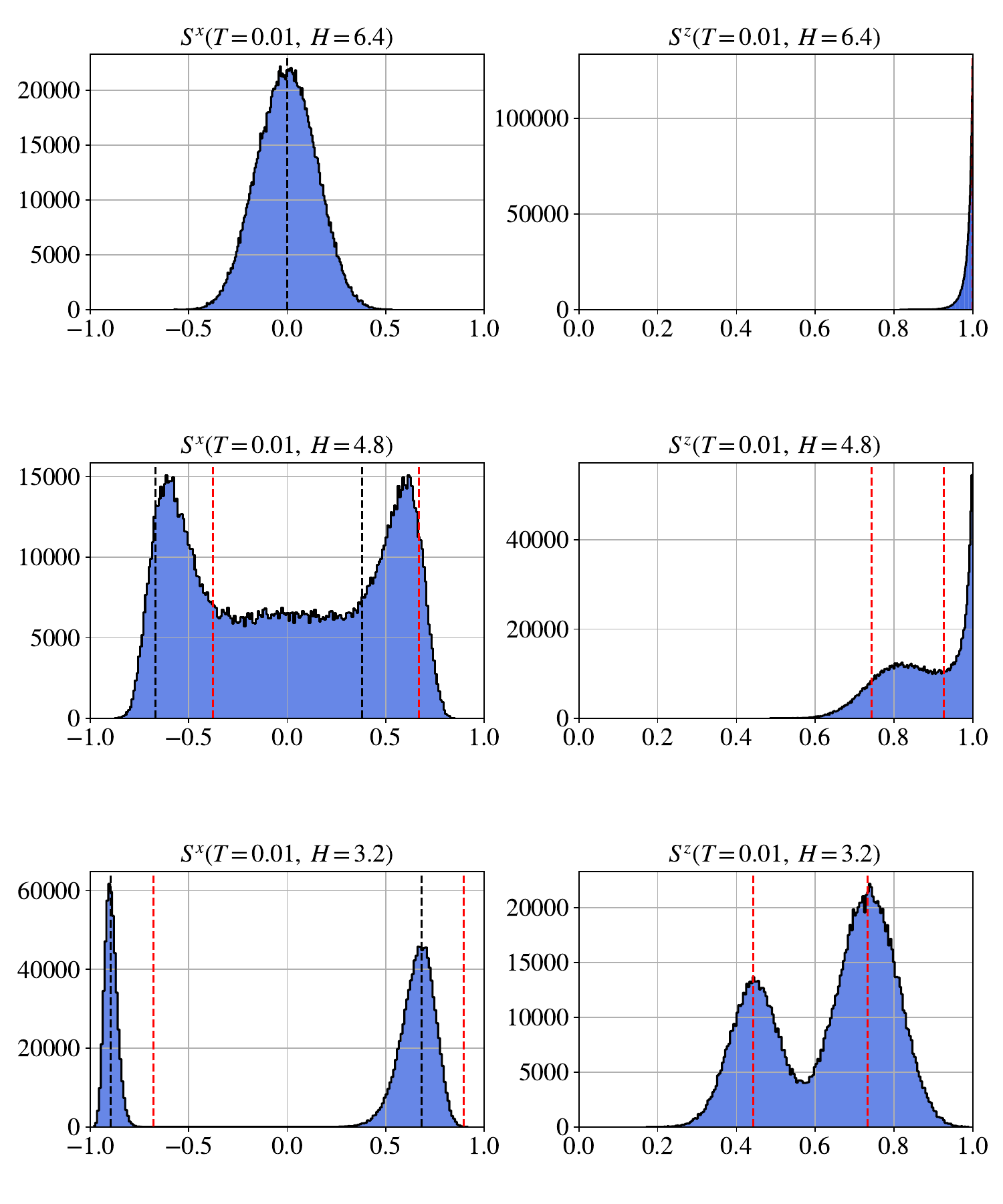}	
    \caption{\label{fig:Distributions}
    Thermal distribution of the $S^x$ and $S^z$ components for a model with $\{J_x,J_y,J_z,J_{xz}\}=\{1,0,0,0\}$ as a function of field for a fixed temperature. Here each row corresponds to a fixed temperature.
    } 
\end{figure}

Figure~\ref{fig:Distributions} illustrates the $x$ and $z$ distributions of the spins for a range of field values at a fixed temperature. In this figure, the vertical dashed lines label the ground-state values of the $S^x$ and $S^z$ components. For the $x$ distribution the two sets of colored lines correspond to the two $\mathbb{Z}_2$ symmetry sectors. At the highest field illustrated, the ground-state corresponds to a fully polarized state, where the $x$ components vanish while the $z$ components become fully saturated. At finite temperatures in the polarized phase (upper panels), the $z$ components of the spins on the Kagome layers fluctuate above the maximal polarization, i.e. $S_i^z=1$. Additionally, the $x$ components of the spins act as fluctuations about the ground-state configuration, resulting in a distribution centered about $S_i^x=0$. As the external field is decreased to $H=4.8$, the system enters the correlated paramagnetic phase (middle panels), where the octupolar antiferromagnetic interaction energetically enforces a 2-in-2-out rule, resulting in a symmetric bimodal distribution of the $x$ components of the spins. The peaks of the $x$ distribution occur  between the possible ground-state values associated with the two $\mathbb{Z}_2$ symmetry sectors. On the other hand, the $z$ components of the spins are distributed according to a bimodal distribution, where one of the peaks of the distribution occurs between the values predicted for the ground-state configuration at the corresponding field. Lastly, as the field is decreased to $H=3.2$, the system enters the fragmented spin liquid phase. In this phase, the $x$ component distribution remains bimodal where the two peaks of the distribution are now centered at values corresponding to one of the two $\mathbb{Z}_2$ symmetry sectors. We note that the $\mathbb{Z}_2$ symmetry-breaking in this phase is also reflected in the missing symmetry of the distribution about the $S^x=0$ axis. In the fragmented spin liquid phase, the $z$ components of the spins are distributed according to a bimodal distribution where the peaks of the distribution match the expected values of the ground-state configurations.

\section{Further Details in $\rm Ce$-based Compounds}
\begin{figure}[ht!]
    \centering
   \includegraphics[width=0.8\columnwidth]{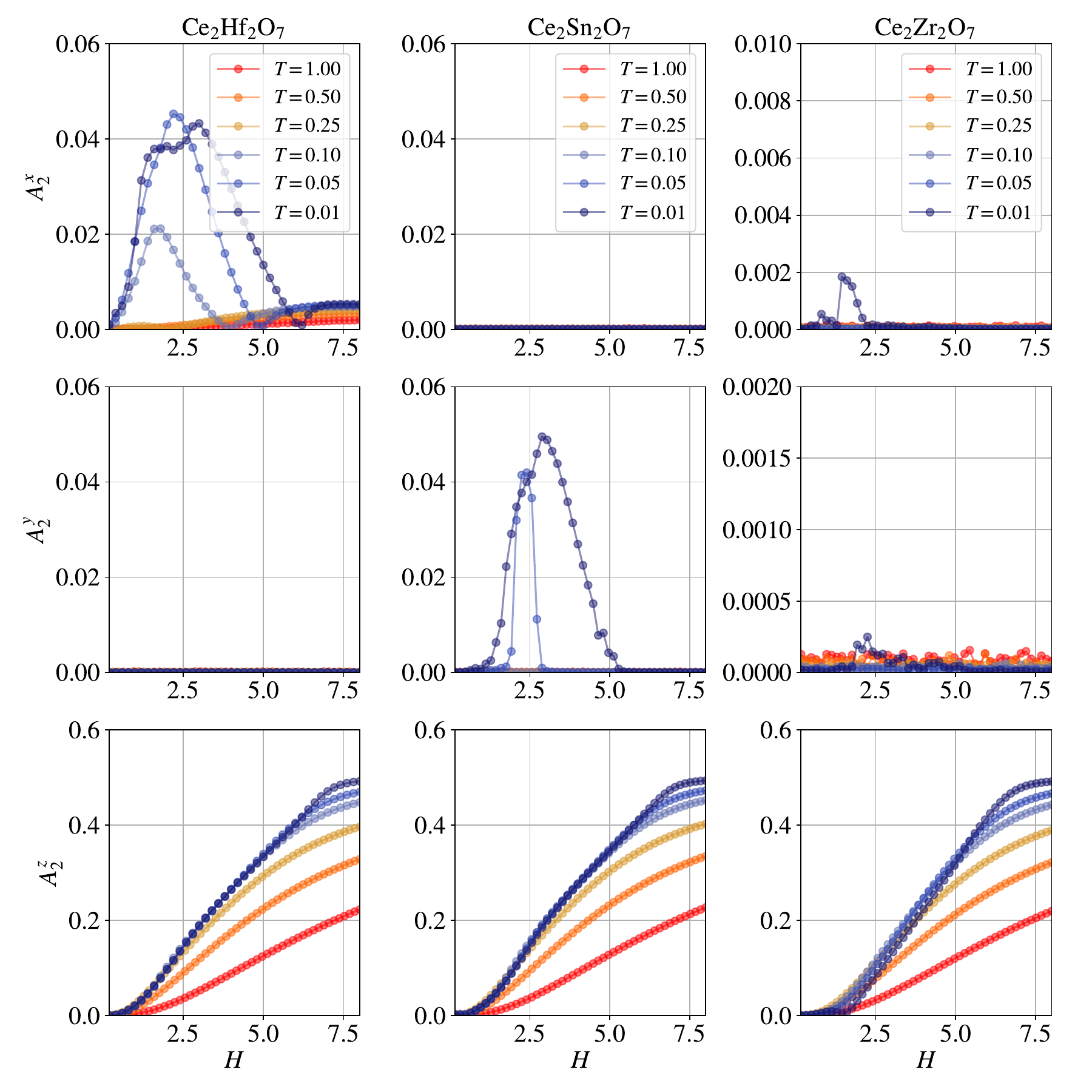}	
    \caption{\label{fig:Materials}
    Prediction of the $A_2^\alpha$ parameters as a function of field and temperature with $\alpha\in \{x,y,z\}$ for distinct fixed temperatures as a function of field. Here, the fist, second and third columns were obtained using the interactions parameters $\{J_x,J_y,J_z,J_{xz}\}=\{1,0.478,0.239,-0.0217\}$~\cite{poree2023dipolaroctupolar}, $\{J_x,J_y,J_z,J_{xz}\}=\{0.208,1,0.208,0\}$~\cite{Sibille2020NatPhys}, and $\{J_x,J_y,J_z,J_{xz}\}=\{1,0.984,0.175,0\}$~\cite{Smith2023PhysRevB}, respectively, where we have taken the highest coupling as the unit of energy.}
\end{figure}

In this section, we present CMC simulations for $\rm Ce_2Hf_2 O_7$, $\rm Ce_2Sn_2 O_7$, and $\rm Ce_2Zr_2 O_7$ for various temperatures as a function of external magnetic field. For these simulations, we have used the interaction parameters presented in Table~\ref{tab:materials} marked by an asterisk and rescale these by the largest coupling. Fig.~\ref{fig:Materials} illustrates the evolution of the $A_2^\alpha$ parameters as a function of field and temperature with $\alpha\in \{x,y,z\}$. We note that for the interaction parameters considered, all of these compounds yield an octupolar spin liquid at low temperatures and vanishing magnetic field.  Under the application of a magnetic field along the  $[111]$  direction, our simulations detect a $\mathbb{Z}_2$ symmetry breaking transition in the octupolar component for which the interaction coupling is dominant. In other words, the $x$ component for $\rm Ce_2Hf_2 O_7$, the $y$ component for $\rm Ce_2 Sn_2 O_7$, and a mix between both components for $\rm Ce_2Zr_2 O_7$  develops a non-vanishing average value signaling the $\mathbb{Z}_2$ symmetry-breaking transition. 

\begin{table}[h!]
    \begin{tabular}{ccccc}
    \toprule
    Material & & \begin{tabular}{cc} Parameters \\ $(J_x, J_y,J_z,J_{xz})$\end{tabular} & \begin{tabular}{cc} single \\ crystal \end{tabular} & comment  \\
    \midrule 
    *$\mathrm{Ce}_2\mathrm{Hf}_2\mathrm{O}_7$~\cite{poree2023dipolaroctupolar} %
    &[Set 1]&$0.046, 0.022,  0.011 , -0.001 $& Yes &  O-SI phase   \\  
    $\mathrm{Ce}_2\mathrm{Hf}_2\mathrm{O}_7$ \cite{smith2025arxiv,chunjiong} 
    &[Set 2]&$0.0432, -0.018,  0.0158 , 0.0166 $& Yes& AIAO phase \\ 
    $\mathrm{Ce}_2\mathrm{Hf}_2\mathrm{O}_7$ \cite{smith2025arxiv,chunjiong} 
    &[Set 3]&$0.0417, 0.021, 0.012, \pm 0.0018 $& Yes  &  O-SI phase   \\ 
    $\mathrm{Ce}_2\mathrm{Sn}_2\mathrm{O}_7$ \cite{YahnePhysRevX.14.011005}& &$-0.027, -0.001, 0.006, 0.026$ & Yes \\
    *$\mathrm{Ce}_2\mathrm{Sn}_2\mathrm{O}_7$ \cite{Sibille2020NatPhys,poree2023fractional} &&$0.01, 0.048, 0.01 , 0$ & No \\
    *$\mathrm{Ce}_2\mathrm{Zr}_2\mathrm{O}_7$ \cite{Smith2023PhysRevB} &&$ 0.063 , 0.062 ,  0.011,  0 $ & Yes\\  
    $\mathrm{Ce}_2\mathrm{Zr}_2\mathrm{O}_7$ \cite{Bhardwaj2022NPJQM} &&$ 0.05 , 0.08,  0.02 ,  0 $ & Yes\\  
    \bottomrule
    \end{tabular}
    \caption{\label{tab:materials}
        Some estimated coupling parameters for DO-QSI candidates, quoted in meV, taken from literature. 
    }
\end{table}

\section{Thermodynamic measurements on distinct parameter sets for $\rm Ce_2Hf_2O_7$}
\begin{figure}[ht!]
    \centering
   \includegraphics[width=0.8\columnwidth]{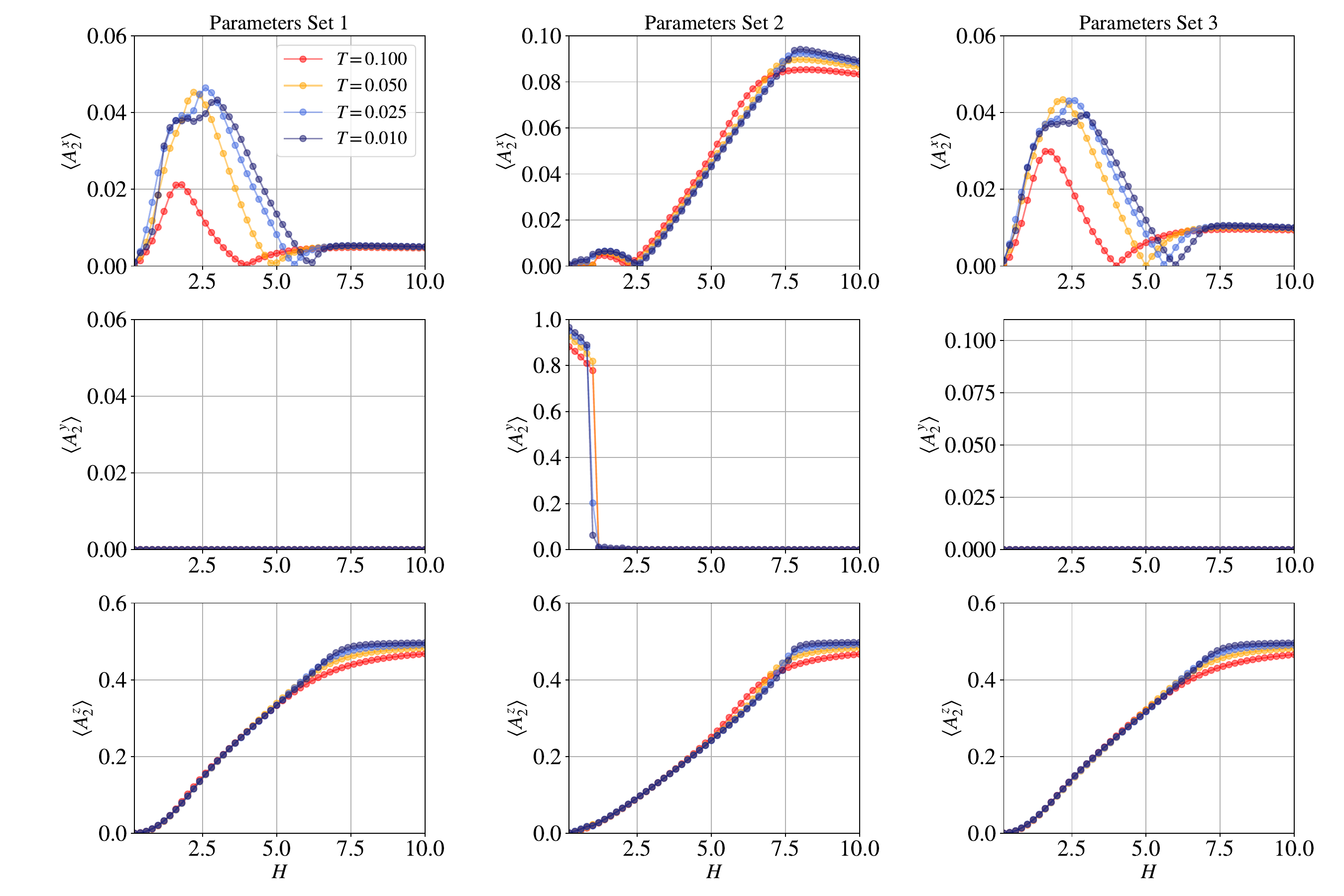}	
    \caption{\label{fig:CeHfO_parameters_A2}
    Prediction of the $A_2^\alpha$ parameters with $\alpha\in\{x,y,z\}$ for distinct parameters for $\rm Ce_2Hf_2O_7$ at fixed temperatures as a function of field. Here, each column correspond to a particular set of parameters given in Table~\ref{tab:materials} as indicated in the title of the first row. }
\end{figure}

\begin{figure}[ht]
    \centering
   \includegraphics[width=0.8\columnwidth]{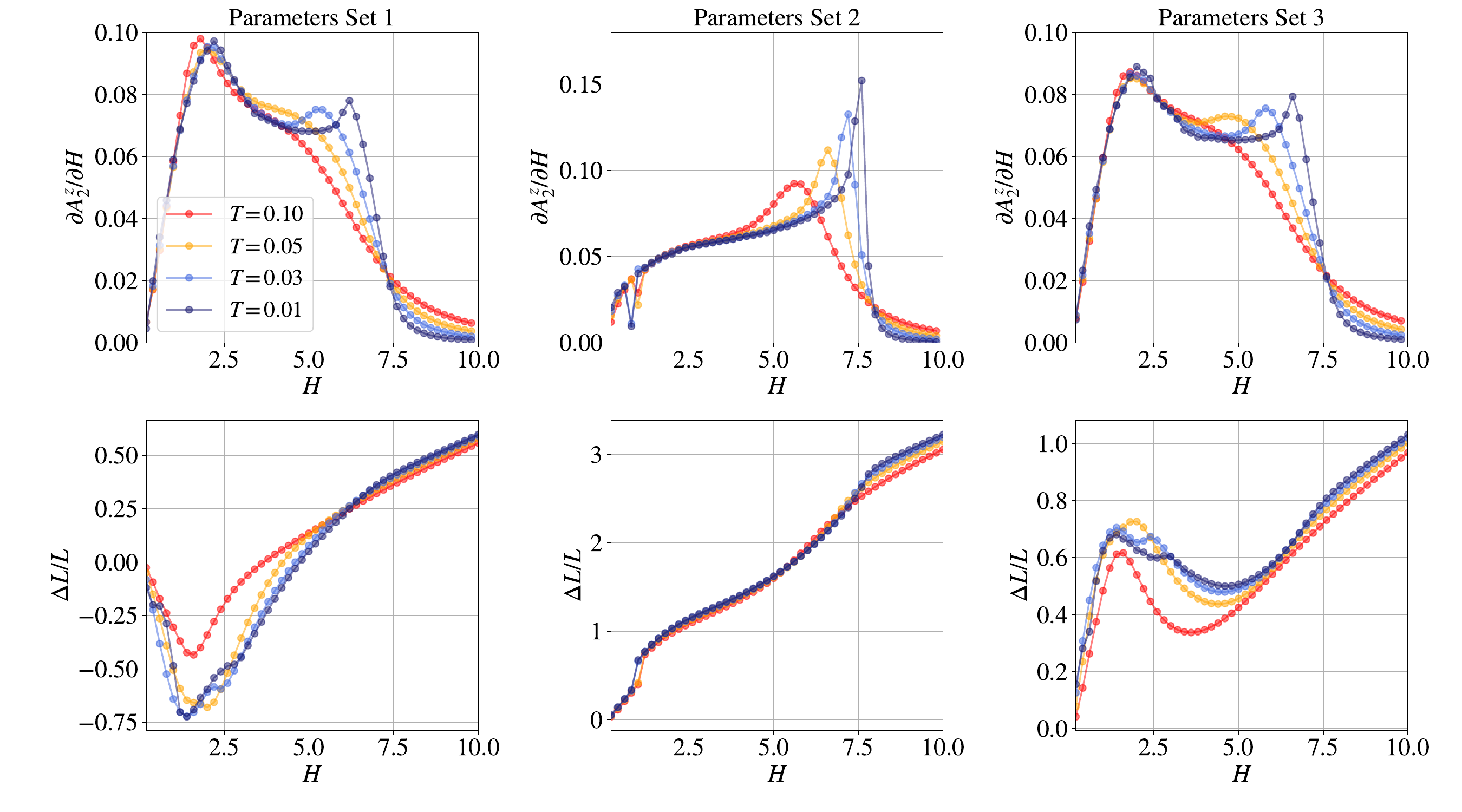}	
    \caption{\label{fig:CeHfO_parameters_thermodynamics}
    Prediction of the $A_2^z$, $\partial A_2^z/\partial H$, and $\Delta L/L$ for distinct parameters for $\rm Ce_2Hf_2O_7$ at fixed temperatures as a function of field. Here, each column correspond to a particular set of parameters given in Table~\ref{tab:materials} as indicated in the title of the first row. 
    }
\end{figure}

\begin{figure}[h!]
    \centering
   \includegraphics[width=0.8\columnwidth]{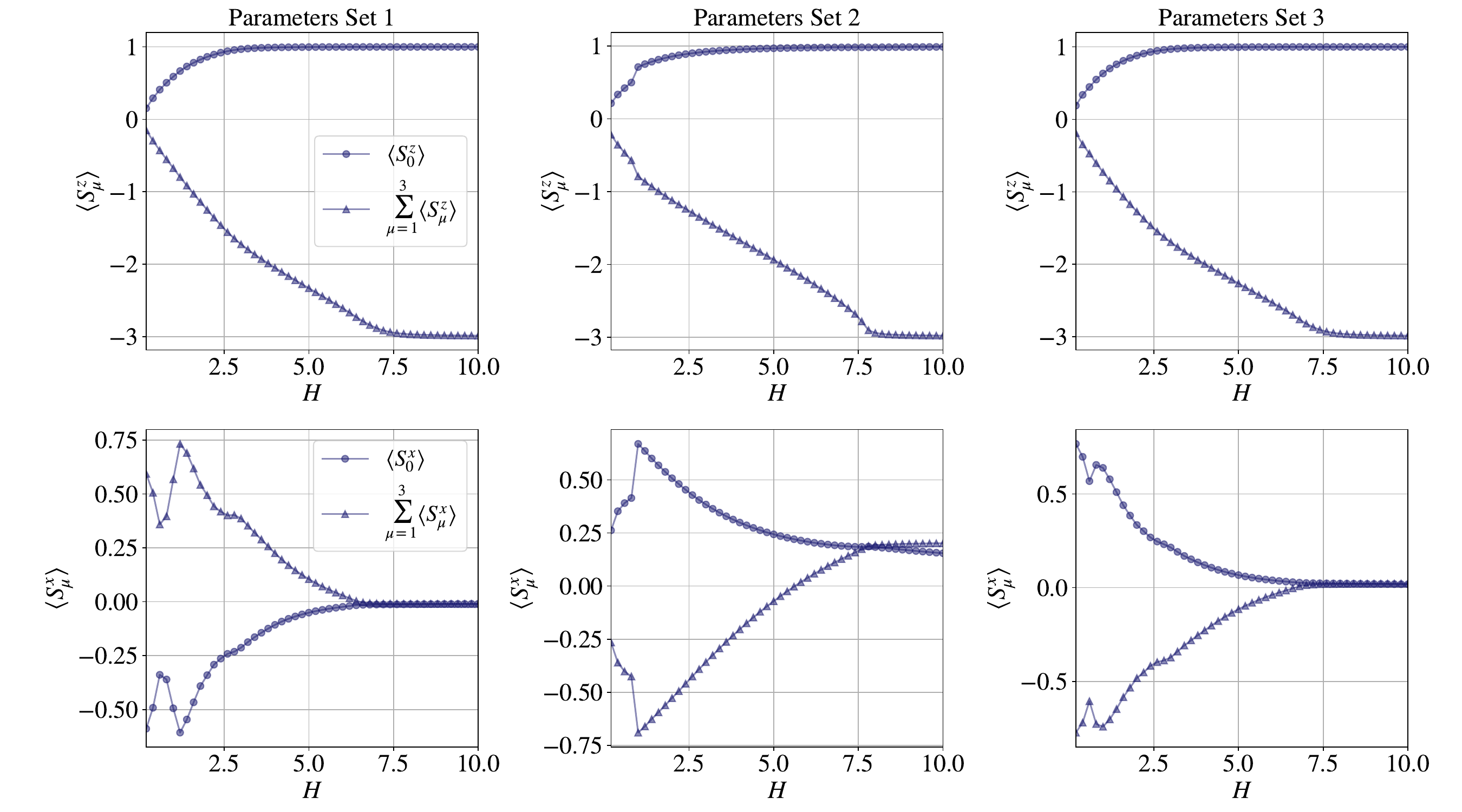}	
    \caption{\label{fig:CeHfO_parameters_spins}
    Thermodynamic average of the spin quantities $\langle S_\mu ^\alpha \rangle $ and $\sum_{\mu=1}^{3} \langle S_\mu ^\alpha \rangle $ with $\alpha\in\{ x,z\}$ as a function of applied magnetic field at fixed temperature $T=0.01$. Here, each column correspond to a particular set of parameters given in Table~\ref{tab:materials} as indicated in the title of the first row. The data illustrated in this figure corresponds to the lowest temperature used in Fig.~\ref{fig:CeHfO_parameters_A2} and ~\ref{fig:CeHfO_parameters_thermodynamics}.}
\end{figure}

\begin{figure}[h!]
    \centering
   \includegraphics[width=0.8\columnwidth]{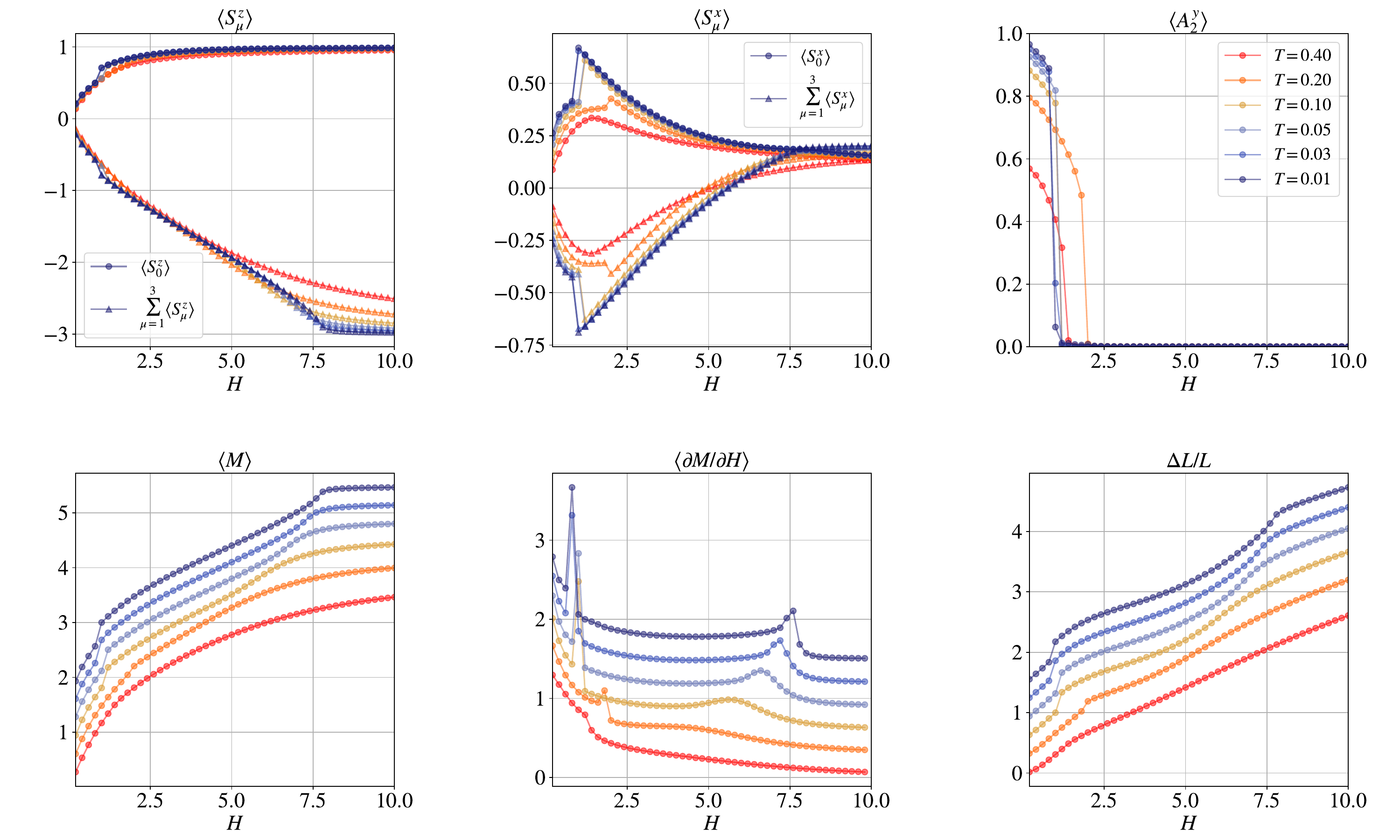}	
    \caption{\label{fig:CeHfO_parameters_experiments}
  Thermodynamic average of the spin quantities $\langle S_\mu ^\alpha \rangle $, $\langle A_2^y\rangle$ order parameter, magnetization $\langle M\rangle$, susceptibility $\langle\partial M/\partial H\rangle$, and magnetostriction $\Delta L/L$ as a function of external magnetic field obtained using the set of parameters $\{J_x,J_y,J_z,J_{xz}\}=\{1,-0.417,0.366,0.384\}$ corresponding to Set 2 of interactions parameters in Table~\ref{tab:materials}, where we have used the $J_x$ coupling as a measure of energy. In this figure, the color of the curves correspond to a particular temperature as labeled in the upper rightmost panel. }
\end{figure}

In this section, we provide the prediction of the $A_2^\alpha$ order parameters with $\alpha\in\{x,y,z\}$, the susceptibility $\partial A_2^z/\partial H$, and the corresponding magnetostriction $\Delta L/L$ measured along the $[111]$ direction, for three different sets of parameters for $\rm Ce_2Hf_2O_7$ illustrated in Table~\ref{tab:materials}. The $A_2^\alpha$ order parameters are illustrated in Fig.~\ref{fig:CeHfO_parameters_A2} where each column correspond to a set of parameters indicated in in Table~\ref{tab:materials}.
The measured thermodynamic quantities as a function of field for various temperatures are shown in Fig.~\ref{fig:CeHfO_parameters_thermodynamics}. The average value of the spins per sublattice used to obtain a measure of the magnetostriction in Fig.~\ref{fig:CeHfO_parameters_thermodynamics} are shown in Fig.~\ref{fig:CeHfO_parameters_spins}. Lastly, in Fig.~\ref{fig:CeHfO_parameters_experiments} we provide additional temperature curves for the parameter set 2~\cite{chunjiong} of the averaged spin quantities $\langle S_\mu ^\alpha \rangle $, averaged order parameter $\langle A_2^y\rangle $, average Magnetization $\langle M\rangle $, susceptibility $\langle \partial M/\partial H \rangle $, and magnetostriction $\Delta L/L$, where the curves of the magnetization, susceptibility and magnetostriction have been vertically displaced to better illustrate their evolution as a function of temperature. In this figure, the unit-less temperatures $T=0.05, 0.1, 0.2,0.4$ approximately correspond to the temperatures $T=100\rm{mk}, 200 \rm {mk}, 400\rm {mk}, 800\rm {mk}$, respectively, where we have used the conversion $\frac{1}{4}\times 0.0432 [\mathrm{meV}]\times \frac{11600 [\mathrm{mK}]}{1[\mathrm{meV}]}$, where the factor $\frac{1}{4}$ accounts for the spin $S=1/2$. For this set of parameters, the system realizes a low-temperature symmetry-breaking $A_2^y$ phase in the absence of external magnetic field, as indicated by the evolution of the $\langle A_2^y\rangle$ parameter. As the external field increases, the $A_2^y$ parameter vanishes and the system enters the fragmented spin liquid phase, as signaled by a sharp peak in the susceptibility $\langle \partial M/\partial H\rangle $, and the small discontinuity in both the magnetization $\langle M\rangle$ and magnetostriction $\Delta L/L$. Lastly, at higher fields, a broad bump in the susceptibility $\langle \partial M/\partial H\rangle $ signals the crossover to the fully polarized phase. At these fields, the magnetization saturates and the magnetostriction becomes linear as a function of magnetic field. 

\section{Coexistence of distinct symmetry sectors}
\begin{figure}[h!]
    \centering
   \includegraphics[width=0.8\columnwidth]{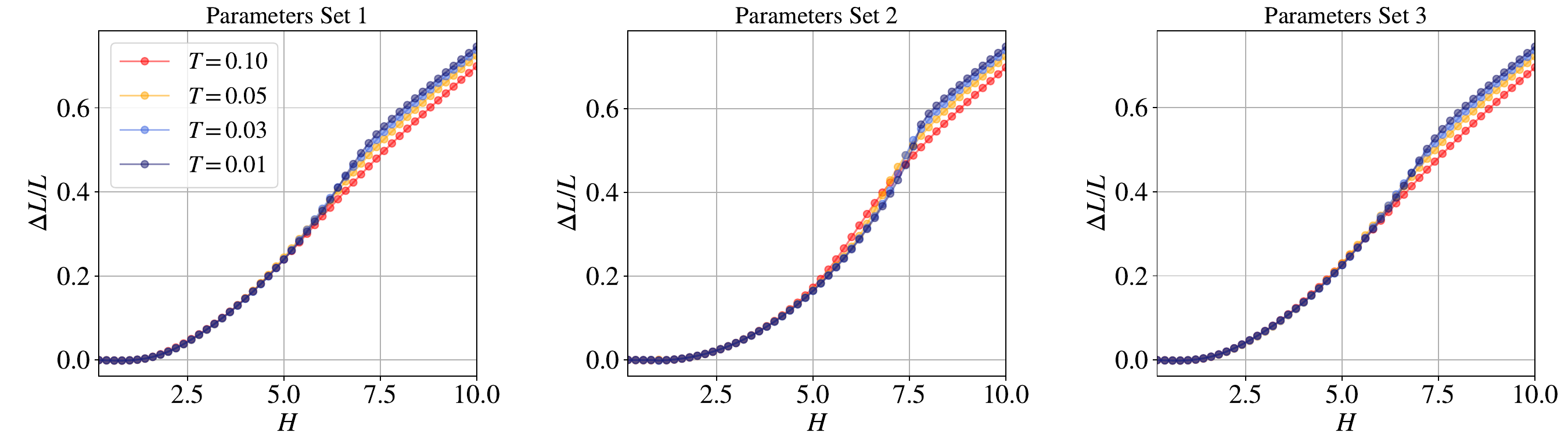}	
    \caption{\label{fig:sectors}
    Magnetostriction obtained for the three sets of parameters considered for $\rm Ce_2Hf_2O_7$ in Table~\ref{tab:materials} where a vanishing average on the local $x$-component of the spins is enforced. These figures should be compared to those shown in Fig.~\ref{fig:CeHfO_parameters_thermodynamics}.
  }
\end{figure}

So far, the simulations carried for the sets of parameters used in Fig.~\ref{fig:CeHfO_parameters_A2},~\ref{fig:CeHfO_parameters_thermodynamics}, and ~\ref{fig:CeHfO_parameters_spins}, result in a single sector of the fragmented spin liquid phase at low temperatures, i.e. one where the gauge-charge associated with the $A_2^x$ parameter is either positive or negative. As discussed in the main text, the simultaneous realization of the two symmetry-sector on a system would yield high-energy domain walls, resulting in a higher energy configuration. However, if this scenario were to take place, the $\mathbb{Z}_2$ symmetry on local $x$-components of the spin, would enforce a vanishing average of the these components, i.e. $\langle S_\mu^x\rangle =0$. In such a case, the magnetostriction in Eq.~\eqref{eq:magnetostriction} would only capture the temperature evolution of the $z$ degrees of freedom. For completeness, in Fig.~\ref{fig:sectors}  we illustrate the magnetostriction for the three parameter sets considered for $\rm Ce_2Hf_2O_7$. For these curves we have enforce $\langle S_\mu^x\rangle =0$, yielding a distinct behavior to the one seen in Fig.~\ref{fig:CeHfO_parameters_thermodynamics}.    

\end{document}